%% file: ms.tex
\documentclass[10pt,apjl]{emulateapj}
\usepackage[dvips]{epsfig}
\usepackage[dvips]{color}

\usepackage{apjfonts}

\begin{document}

\shorttitle{THE TWO POPULATIONS WITH [FE/H] $<$ --3.0}
\shortauthors{Norris et al.}

\newcommand{\cd}     {CD$-38^{\circ}\,245$}
\newcommand{\bd}     {BD$-18^{\circ}\,5550$}
\newcommand{\ito}    {BD$+44^{\circ}\,493$}
\newcommand{\bdbm}   {BD$+18^{\circ}\,5550$}
\newcommand{\hen}    {HE~0107--5240}
\newcommand{\hea}    {HE~1327--2326}
\newcommand{\hej}    {HE~0557--4840}
\newcommand{\hez}    {HE~0057--5959}
\newcommand{\hef}    {HE~1506--0113}
\newcommand{\het}    {HE~2139--5432}
\newcommand{\caffau} {SDSS~J102915+172927}
\newcommand{\teff}   {$T_{\rm eff}$}
\newcommand{\logg}   {$\log g$} 
\newcommand{\kms}    {km s$^{-1}$}
\newcommand{\fei}    {Fe\,{\sc i}}
\newcommand{\feii}   {Fe\,{\sc ii}}

\title{THE MOST METAL-POOR STARS. IV. THE TWO POPULATIONS WITH [FE/H]
  $\la$ --3.0}

\author{
JOHN E. NORRIS\altaffilmark{1},
DAVID YONG\altaffilmark{1},
M. S. BESSELL\altaffilmark{1},
N. CHRISTLIEB\altaffilmark{2},
M. ASPLUND\altaffilmark{1},
GERARD GILMORE\altaffilmark{3},
ROSEMARY F.G. WYSE\altaffilmark{4}, 
TIMOTHY C. BEERS\altaffilmark{5,6}, 
P. S. BARKLEM\altaffilmark{7},
ANNA FREBEL\altaffilmark{8}, AND 
S. G. RYAN\altaffilmark{9}}

\altaffiltext{1}{Research School of Astronomy and Astrophysics, The
  Australian National University, Weston, ACT 2611, Australia;
  jen@mso.anu.edu.au, yong@mso.anu.edu.au,  bessell@mso.anu.edu.au,
  martin@mso.anu.edu.au}

\altaffiltext{2}{Zentrum f\"ur Astronomie der Universit\"at
  Heidelberg, Landessternwarte, K{\"o}nigstuhl 12, D-69117 Heidelberg,
  Germany; n.christlieb@lsw.uni-heidelberg.de}

\altaffiltext{3}{Institute of Astronomy, University of Cambridge,
  Madingley Road, Cambridge CB3 0HA, UK; gil@ast.cam.ac.uk}

\altaffiltext{4}{The Johns Hopkins University, Department of Physics
\& Astronomy, 3900 N.~Charles Street, Baltimore, MD 21218, USA}

\altaffiltext{5}{National Optical Astronomy Observatory, Tucson, AZ
85719, USA}

\altaffiltext{6}{Department of Physics \& Astronomy and JINA: Joint
  Institute for Nuclear Astrophysics, Michigan State University,
  E. Lansing, MI 48824, USA; beers@pa.msu.edu}

\altaffiltext{7}{Department of Physics and Astronomy, Uppsala
 University, Box 515, 75120 Uppsala, Sweden;
 paul.barklem@physics.uu.se}

\altaffiltext{8}{Department of Physics, Massachusetts Institute of
  Technology, Cambridge, MA 02139, USA; afrebel@mit.edu}

\altaffiltext{9}{Centre for Astrophysics Research, School of Physics,
  Astronomy \& Mathematics, University of Hertfordshire, College Lane,
  Hatfield, Hertfordshire, AL10 9AB, UK; s.g.ryan@herts.ac.uk}

\begin{abstract}
We discuss the carbon-normal and carbon-rich populations of Galactic
halo stars having [Fe/H] $\la$ --3.0, utilizing chemical abundances
from high-resolution, high-$S/N$ model-atmosphere analyses.  The
C-rich population represents $\sim28$\% of stars below [Fe/H] = --3.1,
with the present C-rich sample comprising 16 CEMP-no stars, and two
others with [Fe/H] $\sim$ --5.5 and uncertain classification.  The
population is O-rich ([O/Fe] $\ga$ +1.5); the light elements Na, Mg,
and Al are enhanced relative to Fe in half the sample; and for Z $>$
20 (Ca) there is little evidence for enhancements relative to solar
values.  These results are best explained in terms of the admixing and
processing of material from H-burning and He-burning regions as
achieved by nucleosynthesis in zero-heavy-element models in the
literature of ``mixing and fallback'' supernovae (SNe); of rotating,
massive and intermediate mass stars; and of Type II SNe with
relativistic jets.  The available (limited) radial velocities offer
little support for the C-rich stars with [Fe/H] $<$ --3.1 being
binary.  More data are required before one could conclude that
binarity is key to an understanding of this population.  We suggest
that the C-rich and C-normal populations result from two different gas
cooling channels in the very early Universe, of material that formed
the progenitors of the two populations. The first was cooling by
fine-structure line transitions of C\,II and O\,I (to form the C-rich
population); the second, while not well-defined (perhaps dust-induced
cooling?), led to the C-normal group.  In this scenario, the C-rich
population contains the oldest stars currently observed.

\end{abstract}


\section{INTRODUCTION}\label{sec:intro}

Studies of the chemical abundance patterns of the most metal-poor
stars offer insight into the properties and role of the first
generations of stars.  The progenitors of these objects, some likely
to have had zero metallicity, are the first stars to have formed in
the Universe, and may well be responsible for its reionization (e.g.,
\citealt{bromm09}).  The most chemically primitive stars
in the Milky Way hold vital clues concerning the earliest phases
of the formation and evolution of the Galaxy.

Extensive observation of metal-poor candidates in the HK survey
\citep{beers85, beers92}, the Hamburg/ESO Survey (HES; \citealt{hes,
  christlieb08}), the Sloan Digital Sky Survey (SDSS; \citealt{sdss})
and the SEGUE survey \citep{segue} has greatly increased the sample of
extremely metal-poor stars ([Fe/H] $<$ $-$3.0)\footnote{For element X,
  [X/H] = log(N(X)/N(H))$_{\star}$ -- log(N(X)/N(H)$_{\odot}$.}.
Subsequent chemical abundance analyses of the brightest of these have
revealed that, in addition to a population of apparently ``normal''
metal-poor stars (those having well-defined trends for most elements),
there exist chemically peculiar stars (those with strong enhancements
or deficiencies of particular elements) (see e.g.,
\citealt{mcwilliam95, ryan96, norris01, johnson02, cayrel04, cohen08,
  lai08, yong12}).  With the discovery and analysis of
non-``normal'' stars at lowest [Fe/H], different classes of objects
are being defined which are permitting exploration of the nature and
frequency of the progenitor stars that were responsible for this rich
diversity, accompanied by insight into chemical enrichment of the
Universe at the earliest times.

The best known type of chemically anomalous object at very low
metallicity is the carbon-enhanced metal-poor (CEMP) class
(\citealt{beers05}), which comprises a large fraction ($\sim10-20\%$)
of metal-poor stars below [Fe/H] = --2.0.  As defined by
\citet{beers05} (and as discussed below in Section~\ref{sec:catalog}),
the CEMP class itself has several distinct subclasses.  This
diversity of chemical properties is not, however, confined to
carbon-rich stars.  Rarer examples of individual chemically unusual
low-[Fe/H], non-CEMP, stars include (1) the Mg-enhanced metal-poor
star BS~16934-002, with [Fe/H] = --2.7, [Si/Fe] = +0.44, and [Ca/Fe] =
+0.35, but [Mg/Fe] = +1.23 \citep{aoki07b}; (2) the $\alpha$-element
challenged HE~1424--0241, with [Fe/H] = --4.0 and [Mg/Fe] = +0.44, but
[Si/Fe] = $-$1.01 and [Ca/Fe] = --0.44 \citep{cohen07}; and (3) the
$\alpha$-element ambivalent SDSS J234723.64+010833.4, with [Fe/H] =
--3.17 and [Mg/Fe] = $-$0.10, but [Ca/Fe] = +1.11 \citep{lai09}. In
Paper~II of the present series \citep{yong12} we reported a
homogeneous chemical analysis of 190 metal-poor stars, and presented
chemical abundances for some 16 elements.  In that sample there were
109 stars for which we were able to determine the CEMP/C-normal
status, and which are C-normal (i.e., stars with [C/Fe] $<$
+0.7)\footnote{We caution the reader that this does not represent a
  full inventory, since it excludes C-normal stars in which C was not
  detected and for which the upper limit on [C/Fe] was greater than
  +0.7.}.  For elements in the range Na--Ni, we determined the
incidence of anomalous abundances relative to Fe (where anomalous is
taken to mean [X/Fe] different from the average ``normal'' star value
by more than $\pm$0.5 dex).  We found that 21 $\pm$ 5 \% of stars were
anomalous with respect to one element, while 4 $\pm$ 2 \% were
anomalous with respect to at least two.

Given the small number of stars presently known at extremely low
[Fe/H], the identification of just a handful with similar chemical
properties can not only define a class of stars, but also reveal that
what was originally regarded as a rare and peculiar object may indeed
represent a more substantial fraction of the population. For example,
since the discovery and analysis of the highly r-process-element
enhanced star CS~22892--052 \citep{mcwilliam95,sneden96}, several
additional examples of this class have been identified (see
\citealt{sneden08} and references therein).  Another example more
pertinent to the present investigation is the case of the CEMP stars
CS~22949-037 \citep{mcwilliam95,norris01,depagne02} and CS~29498-043
\citep{aoki02b}, both with no enhancement of the heavy neutron-capture
elements, but which \citet{aoki02b} identified as a subclass of the
CEMP stars which also has enhanced Mg and Si.

This is the fourth paper in the our series on the discovery and
analysis of the most metal-poor stars.  Here we focus on the detailed
chemical abundance patterns of the C-rich stars having [Fe/H] $\la$
--3.0 (many with large enhancements of some or all of Na, Mg, Al, and
Si), and what they have to tell us about the origin of the remarkable
increase of carbon richness, not only in frequency but also in degree,
as [Fe/H] decreases.  In Section~\ref{sec:catalog} we present a sample
of C-rich stars with [Fe/H] $\la$ --3.0 (excluding those having large
heavy-neutron-capture-element enhancements), based principally on our
homogeneous chemical analysis of Paper
II. Sections~\ref{sec:abundances} -- \ref{sec:kinematics} discuss the
chemical abundances and kinematics of this sample in comparison with
C-normal stars in the same metallicity range.  In
Section~\ref{sec:discussion} we then consider the relevance of various
models that have been proposed to explain the origin of anomalous
abundances in the early Universe.  We shall argue for the existence of
two principal channels of cooling and chemical enrichment to explain
the C-rich and C-normal populations observed at lowest Fe abundance,
[Fe/H] $\la$ --3.0.

\section{A SAMPLE OF C-RICH (CEMP-NO AND TWO HYPER METAL-POOR) STARS WITH [FE/H] $\la$ --3.0}\label{sec:catalog}

\citet{aoki10} demonstrated that, for [Fe/H] $<$ --3.0, the large
majority ($\sim90\%$) of CEMP stars belong to the CEMP-no subclass,
and it is these objects that will concern us here.  \citet{beers05}
define a CEMP-no star as one having [C/Fe] $>$ +1.0 and [Ba/Fe] $<$
0.0.  The other CEMP subclasses, defined by these authors, all of
which have [C/Fe] $>$ +1.0, are: (i) CEMP-r -- [Eu/Fe] $>$ +1.0; (ii)
CEMP-s -- [Ba/Fe] $>$ +1.0 and [Ba/Eu] $>$ + 0.5; and (iii) CEMP r/s
-- 0.0 $<$ [Ba/Eu] $<$ +0.5.  More recently, \citet{aoki07} have
suggested a slightly modified [C/Fe] criterion, based on extensive
high-resolution, high-$S/N$ abundance analysis, also taking into
account putative stellar evolution effects at highest luminosity.
Based on their Figure~4, their definition is (i) [C/Fe] $\ge$ +0.7,
for $\log (L/L_\odot) \le 2.3$ and (ii) [C/Fe] $\ge$ +3.0 $-$ $\log
(L/L_\odot)$, for $\log (L/L_\odot) > 2.3$. In what follows, we shall
somewhat more conservatively adopt [C/Fe] $\ge$ +0.70 for all values
of $\log(L/L_{\odot}$), since it is not clear to us that criterion
(ii) is sufficiently justified, based on the data in our
sample\footnote{Our concern is based, for example, on luminous stars
  such as {\bdbm} in Table 6 of Paper II, which has {\teff} = 4560\,K,
  {\logg} = 0.8, [Fe/H] = --3.2, [C/Fe] = --0.02, and quite normal
  relative abundances of all elements.  It has $\log(L/L_{\odot}$) =
  3.1, and from criterion (ii) a limiting value for C-rich status of
  [C/Fe] = --0.1, above which stars are accepted as CEMP.  We are
  reluctant to accept objects such as this as CEMP stars.}.

In order to compare the intrinsic abundance patterns of C-rich and
C-normal stars at lowest [Fe/H] values, we begin by selecting C-rich
stars with [Fe/H] $\la$ --2.5, excluding stars of the CEMP-r, CEMP-r/s
and CEMP-s subclasses.  In subsequent discussion we shall then
restrict our consideration essentially to the subset of CEMP-no stars
having [Fe/H] $<$ --3.0.  Insofar as the abundance characteristics of
the CEMP-s, and presumably the CEMP-r/s stars, are driven in large
part by binarity and mass transfer from an asymptotic giant branch
(AGB) star onto the star now being observed (see, e.g.,
\citealt{lucatello05b}), the case for removal of these stars is clear.
That is, abundance patterns driven by extrinsic factors severely
compromise interpretation of the chemical abundances with which these
objects formed.  The exclusion of the CEMP-r group is also
well-justified, given its large r-process-element enhancements.  That
said, given the intrinsic differences among the CEMP subclasses, our
selection processes described below yield no CEMP-r, CEMP-r/s, or
CEMP-s stars with [Fe/H] $<$ --3.1\footnote{We do not claim that such
  stars do not exist.  Rather, our position is that they are rare and
  absent from our sample of C-rich stars with [Fe/H] $<$ --3.1, and
  not important in the present context.}.

Our sample of C-rich stars contains essentially only members of the
CEMP-no subclass, and was selected from three sources, as follows.  We
began with Paper II, in which we have presented chemical abundances,
determined using 1D, local thermodynamic equilibrium (LTE)
model-atmosphere techniques, for a sample of 38 metal-poor stars
having [Fe/H] $\la$ --3.0, of which 34 were newly discovered.  Among
these objects, some nine are CEMP-no stars\footnote{The identification
  of 53327-2044-515, which has [Ba/Fe] $<$ +0.34, as CEMP-no is not
  robust.  Even if it has [Ba/Fe] = +0.34, however, we would regard it
  as being closely related to the class.}.  Second, in Paper~II, we
also determined abundances for a further 152 metal-poor stars in the
literature for which equivalent widths were published and the
atmospheric parameters {\teff} and {\logg} could be reliably
determined from publicly available data. For these, we re-determined
abundances using the same techniques as for our sample of 38 stars.
This literature collection contains 12 C-rich stars of interest for
the present investigation.  Ten of them are CEMP-no stars, while the
other two have [Fe/H] $\sim$ --5.5, [C/Fe] $\sim$ +4, but only [Ba/Fe]
limits, which precludes determination of their CEMP status.  The
combined sample of 190 stars in Paper II comprises chemical abundances
based on high-resolution ($R$ $\sim40,000$), high $S/N$ material,
homogeneously analyzed.  Third, we adopted abundances from the
literature for an additional two CEMP-no stars, {\ito} \citep{ito09}
and Segue~1-7 \citep{norris10b}, that were not considered in Paper II.

Table~\ref{tab:basic_cempno} presents our resulting catalog of the
collective sample of 23 C-rich stars (excluding the CEMP-r, CEMP-r/s
and CEMP-s subclasses) having [Fe/H] $<$ --2.5.  Columns (1) -- (6)
present star name and coordinates, together with the atmospheric
parameters {\teff}, {\logg}, and [Fe/H], respectively, Columns (7) --
(9) contain [C/Fe], [Sr/Fe], and [Ba/Fe], which inform the
identification of most of these objects as CEMP-no stars.  As noted
above, two stars in Table~\ref{tab:basic_cempno} cannot be classified
as CEMP-no for the technical reason that only limits are available for
their barium abundances.  These are {\hen} and {\hea}, the two most
Fe-poor stars currently known, for which [Ba/Fe] $<$ +0.82 and $<$
+1.46, respectively.  This uncertainty notwithstanding, both have
[C/Fe] $\sim$ +4, and are among the most C-rich Fe-poor stars known.
Further, as we shall see below, {\hea} has large enhancements of Na,
Mg, and Al, which is characteristic of half of the C-rich stars with
[Fe/H] $<$ --3.1.  The final column contains source material relevant
to discovery and identification of CEMP-no stars.  We note that
inspection of the table reveals that some seven, or 30\%, of the stars
have {\teff} $>$ 5500\,K, and may be regarded as near turnoff or
subgiant objects, while the remainder are red giants.

\input{tab1}

The above source material also leads to identification of members of
the other CEMP subclasses.  For completeness and the interest of
others, we present in Table~\ref{tab:basic_cemprs} the corresponding
catalog of 26 additional CEMP stars included in the selections of
Paper II that have [Ba/Fe] $>$ 0.0, where the columns have the same
content as in Table~\ref{tab:basic_cempno}.  We regard these as
non-CEMP-no stars and, by process of elimination, members of the
\citet{beers05} CEMP subclasses r, r/s, and s.  The reader will note
that all stars in this sample have [Fe/H] $>$ --3.1.  In considerable
contrast, the techniques of Paper II were strongly biased towards the
recognition and analysis of stars having {Fe/H] $\la$ --3.0.  We
emphasize that Table~\ref{tab:basic_cemprs} is thus potentially
seriously incomplete for abundances larger than this limit.

\input{tab2}

Finally, in Table~\ref{tab:barklem}, we supplement the material
presented above with data for CEMP stars from the work of
\citet{barklem05}\footnote{http://www.astro.uu.se/$\sim$barklem/},
which we did not analyze in Paper II.  This table contains information
for an additional six CEMP-no stars and 14 from the CEMP-r, -r/s, and
-s subclasses.

\input{tab3}

The lower panel of Figure~\ref{fig:cfe_vs_feh} presents the data from
Tables~\ref{tab:basic_cempno}, \ref{tab:basic_cemprs}, and
\ref{tab:barklem} in the ([C/Fe], [Fe/H]) -- plane\footnote{We draw
  the reader's attention to the fact that for two stars,
  53327-2044-515 and HE~1201--1512, we present both subgiant and dwarf
  abundance solutions from Paper II in Table~\ref{tab:basic_cempno},
  while in Figure~\ref{fig:cfe_vs_feh} (and all other figures in this
  paper) we plot them each only once, adopting their average values.},
where the values for the C-rich stars in Table~\ref{tab:basic_cempno}
and the CEMP-no stars in Table~\ref{tab:barklem} are plotted as red
crossed circles (first sample), and the CEMP-r, -r/s, and -s stars
from Tables~\ref{tab:basic_cemprs} and \ref{tab:barklem} (second
sample) are presented as blue dotted circles. The small filled circles
represent the C-normal stars in Paper II that have carbon detections,
while the large filled circle shows the upper limit for the ultra
metal-poor dwarf {\caffau} \citep{caffau11, caffau12}.  The upper
panel of the figure contains the generalized histograms of the two
samples and confirms the metallicity ([Fe/H]) distribution difference
between the CEMP-s and CEMP-no subclasses documented by
\citet{aoki10}.  We note in particular that while CEMP-no stars are
found at all metallicities below [Fe/H] $\sim$ --2.0 in our samples
(Tables~\ref{tab:basic_cempno} and \ref{tab:basic_cemprs}), there are
no CEMP-r, -r/s, or -s stars with [Fe/H] $<$ --3.1.

\begin{figure}[!tbp]
\begin{center}
\includegraphics[width=7.5cm,angle=0]{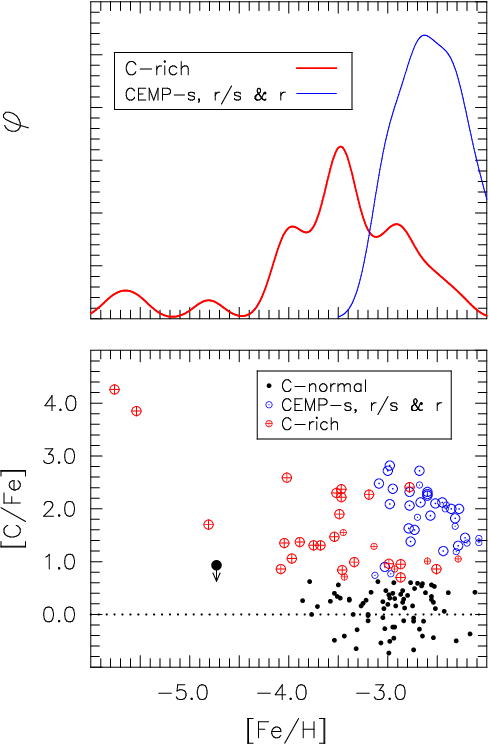}

\caption{\label{fig:cfe_vs_feh} \small (Lower panel) [C/Fe] vs. [Fe/H]
  for the C-rich stars (CEMP-no stars and two with [Fe/H] $\sim$ --5.5)
  (large crossed circles) and CEMP-r, r/s, -s stars (large dotted
  circles) in Tables~\ref{tab:basic_cempno} and
  \ref{tab:basic_cemprs}, respectively.  Smaller symbols are used for
  the data of \citet{barklem05} in Table~\ref{tab:barklem}.  The large
  filled circle represents the ultra metal-poor, C-normal, star
  {\caffau}, while the small filled circles stand for C-normal stars
  in Paper II for which carbon abundances are available. (Upper panel)
  Generalized histograms (with gaussian kernel $\sigma=0.15$) of
  CEMP-no plus two C-rich stars with [Fe/H] $\sim$ --5.0 (thick line) and
  CEMP-s, -r/s and -r stars (thin line). }

\end{center}
\end{figure}

The critic might note that, for [Fe/H] $\la$ --3.0, one cannot make a
strong case from Figure~\ref{fig:cfe_vs_feh} that the C-rich and
C-normal stars represent populations having distinct carbon
abundances, and suggest rather a continuum of [C/Fe] values to which
we have applied an arbitrary dividing line.  We would agree that one
cannot make the former case, and reply that the data do not
necessarily support either position.  We make two points.  First, the
suggestion of a continuous distribution of [C/Fe] in
Figure~\ref{fig:cfe_vs_feh} is in some contrast to what one sees in
Figure 3 ([C/Fe] vs. [Fe/H]) and Figure 5 ([C/Fe] vs
log(L/L$_{\odot}$) of \citet{aoki07}, where clear separations are
evident.  This difference could result at least in part from the fact
that our carbon abundances comprise a more heterogeneously determined
data set than that of Aoki et al., leading to an apparent overlap of
two distinct [C/Fe] distributions.  Second, however, and more to the
point of our approach, as one progresses from [Fe/H] = --3.0 to lower
values of [Fe/H], the relative incidence of C-rich stars appears to
increase.  If the distribution of [C/Fe] is continuous at a given
value of [Fe/H], our Figure~\ref{fig:cfe_vs_feh} suggests that the
form of that distribution changes towards one favoring larger values
of [C/Fe], as [Fe/H] decreases.  It is that change in the form of the
[C/Fe] distribution, and its origins, that we seek to understand.

It is also of interest to estimate the C-rich fraction in the present
sample.  Given that many dwarfs only have [C/Fe] limits that are
significantly greater than +0.7, we consider only stars with {\teff}
$\le$ 5510\,K, which we regard as giants.  Then, below [Fe/H] = --3.1,
there are 13 C-rich giants in Table~\ref{tab:basic_cempno}, of which
11 were included in the stars analyzed of Paper II.  In that sample,
40 stars were C-normal (with [C/Fe] $\le+0.7$), all with no limiting
value greater than +0.7.  This leads to a C-rich fraction of 28 $\pm$
9\%.  The error estimate is based solely on the propagation of errors
in the observed numbers of stars involved in computing the fraction;
it does not include the effect of errors in abundance determinations
or selection biases (presumably towards preferential inclusion of
C-rich giants) that may be inherent in the collective sample that was
analyzed, principally, in Paper II.

In what follows we shall concentrate exclusively on the C-rich stars
of Table~\ref{tab:basic_cempno} that have [Fe/H] $<$ --3.0, which we
shall refer to as ``the C-rich stars''.  In summary, this sample
comprises 18 objects -- 16 from the CEMP-no subclass, together with
two other objects discussed above that have [Fe/H] $\sim$ --5.5, but
only upper limits to their barium abundances.

\section{THE ABUNDANCE PATTERNS OF THE C-RICH STARS WITH [FE/H] $\la-3.0$}\label{sec:abundances}
 
Table~\ref{tab:abundances} presents [Fe/H], relative abundances
([X/Fe]) for C (repeated for convenience), N, O, Na, Mg, Al, and Ca --
Ni (excluding V), together with $^{12}$C/$^{13}$C, for the stars in
Table~\ref{tab:basic_cempno}.  Relevant source material is also
included in the final column of the table.  We emphasize that all of
the abundances in the table are based on high-resolution, high-$S/N$
data.  That said, we also recall that all of these values were
determined using one-dimensional (1D), LTE, model-atmosphere analyses.
It would obviously be preferable to have results based on 3D, non-LTE
techniques.  This is, however, beyond the scope of the present
investigation.  In what follows, we shall also present results for
C-normal stars obtained using 1D, LTE analysis, which permit a
differential comparison, at given [Fe/H], between the C-rich and
C-normal populations with [Fe/H] $\la-$3.0.

\input{tab4}

\subsection{CNO Abundances}

The relative CNO abundances of the C-rich stars with [Fe/H] $<$ --3.0
are presented in Figure~\ref{fig:cno}, as a function of [Fe/H] and
[C/Fe], where the data from Table~\ref{tab:basic_cempno} are plotted
as square and star symbols, for stars having [Fe/H] $\le-4.3$ and
--4.3 $<$ [Fe/H] $\le-3.0$, respectively.  For comparison purposes,
abundances for C-normal (i.e., non-CEMP) red giants from the works of
\citet{cayrel04} and \citet{spite05}, together with those of the ultra
metal-poor main-sequence dwarf, {\caffau}, from \citet{caffau11,
  caffau12} (at [Fe/H] = --4.7) are also plotted, as circles.  In the
left panels, relative abundances [C/Fe], [N/Fe], and [O/Fe] are
presented as a function of [Fe/H], where dotted lines in the figure
represent solar abundance ratios for the ordinate. Also shown in
Figure~\ref{fig:cno} (and in Figures~\ref{fig:c1213} -- ~\ref{fig:omg}
that follow) are representative error bars for the C-rich stars, based
principally on error estimates presented in Paper II, together with
sources cited in that paper.  From these three panels we note (i)
below [Fe/H] $\la$ --4.3, three of the four stars are carbon rich and
(ii) carbon-richness is generally accompanied by both nitrogen and
oxygen enrichment.  This is also clear in the two panels at bottom
right, which show strong correlations of both [N/Fe] and [O/Fe] with
[C/Fe].  For the C-rich stars, the large carbon values reflect their
selection criteria.  For oxygen, on the other hand, the accompanying
extreme enhancements of [O/Fe] are ubiquitous, remarkable, and not the
result of any selection effect\footnote{We note for completeness that
  estimates of the oxygen abundance are not available for nine of the
  18 stars having [Fe/H] $<$ --3.0 in Table~\ref{tab:abundances}.
  (Six of them do have [O/Fe] limits that are not inconsistent with
  the trend seen in the bottom right panel of Figure~\ref{fig:cno}.)
  While, in part, this may be due to the greater difficulty of
  measuring the abundance of O in comparison with that of C, it could
  in principle be due to lower values of [O/Fe] than might be expected
  from the correlation seen in the figure. Further investigation is
  necessary to constrain this possibility.}.

\begin{figure*}[!tbp]
\begin{center}

\includegraphics[width=12.0cm,angle=0]{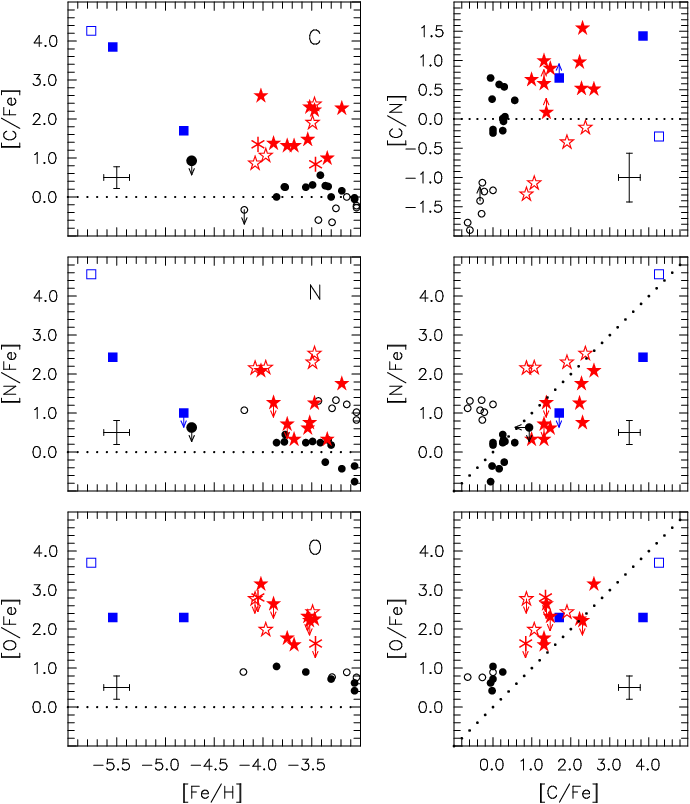}
  
\caption{\label{fig:cno} \small Left column: relative abundances
  [C/Fe], [N/Fe], and [O/Fe], as a function of [Fe/H], for the C-rich
  (CEMP-no and two hyper metal-poor) stars of
  Table~\ref{tab:abundances} (square and star symbols) and C-normal
  stars (small and large circles, from \citealp{cayrel04}
    and \citealp{caffau11}, respectively).  Squares and stars
    represent objects with [Fe/H] $<$ --4.5 and [Fe/H] $>$ --4.5,
    respectively.  Right column: [C/N], [N/Fe] and [O/Fe] vs. [C/Fe].
    Filled and open square and star symbols in all panels are used for
    stars that have [C/N] greater or less than zero (the solar value);
    for stars with [Fe/H] $>$ --4.5, filled and open circles refer to
    ``unmixed'' and ``mixed'' stars; and asterisks represent stars
  that have no estimate of N abundance.  See text for source
  information and discussion.}

\end{center}
\end{figure*}

Before discussing the C-rich stars further, we comment on the C-normal
red giant stars -- in particular the spreads in C and N seen in
Figure~\ref{fig:cno} at a given [Fe/H], and the clear separation into
two groups.  \citet{spite05}, to whom we refer the reader, explain
these in terms of internal mixing effects within the stars currently
being observed, during their evolution on the red giant branch (RGB).
The C-normal giants in Figure~\ref{fig:cno}, represented by small open
and filled black circles, are described by \citet{spite05} as
``mixed'' and ``unmixed'', respectively.  They argue that the mixed
stars have reduced carbon and enhanced nitrogen abundances as the
result of internal CNO processing, together with subsequent mixing of
the processed material to the stellar surface during RGB evolution.  A
second point to note is that the range in [C/Fe] among the C-normal
stars is considerably smaller that than seen among the C-rich objects.

As may be seen in the middle right panel of the Figure~\ref{fig:cno},
extreme enhancements of nitrogen exist among the C-rich stars.  In the
range +0.8 $<$ [C/Fe] $<$ +1.3, the relative nitrogen abundance of
these stars exhibits a large range, +0.3 $<$ [N/Fe] $<$ +2.1,
suggestive perhaps of the existence of variable degrees of processing
via the CN cycle.

\subsection{[C/N] and $^{12}$C/$^{13}$C }\label{sec:c1213}

In the upper right panel of Figure~\ref{fig:cno} we plot [C/N]
vs. [C/Fe], where the large range of [C/N] among the C-normal stars is
clearly seen.  A large separation appears to be present not only
between the ``mixed'' and ``unmixed'' C-normal stars, but also among
the C-rich stars, and we use open and filled blue squares, on the one
hand, and open and filled red star symbols, on the other, to designate
C-rich stars that lie below and above [C/N] = 0.0, respectively. Our
choice of filled and open symbols was made to permit the reader to
appreciate the degree of CN processing that may have been experienced
by the material in the star's outer layers.  (For those stars in
Table~\ref{tab:abundances} having no estimate of [N/Fe], here and in
what follows, we use the asterisk symbol.) Closer inspection of
Table~\ref{tab:abundances} and Figure~\ref{fig:cno} shows that the
\citet{spite05} intrinsic ``mixing'' explanation for the CN patterns
in the normal stars cannot be the full explanation for the patterns of
the C-rich objects: the most iron-poor star {\hea}, with [C/Fe] = +4.3
and [C/N] = --0.3, is a near-main-sequence-turnoff subgiant ({\teff} =
6180\,K and {\logg} = 3.7; \citealp{frebel08}), which has presumably
not yet experienced the mixing of CN processed material from its
interior into its outer layers (as was invoked by \citealt{spite05} to
explain the lower values of [C/N] found in red giants).

In the three panels of Figure~\ref{fig:c1213} we plot
$^{12}$C/$^{13}$C as a function of [Fe/H], [C/Fe], and [C/N], for
stars having [Fe/H] $<$ --3.0, using the data in
Table~\ref{tab:abundances}.  Bearing in mind the caveat that 10 of the
15 stars represented in the figure have only lower limits, we note
that in the top panel one sees perhaps the suggestion of a positive
correlation between $^{12}$C/$^{13}$C and [C/N], in the sense that
would be expected from the processing of hydrogen and carbon in the
CN-cycle.  The large values of [C/Fe] seen in Figure~\ref{fig:c1213},
however, suggest that, if this were the case, one would require two
processes, involving not only the CN-cycle, but helium burning as
well.  We shall return to this point in Section~\ref{sec:discussion}.

\begin{figure}[!tbp]
\begin{center}
\includegraphics[width=7.0cm,angle=0]{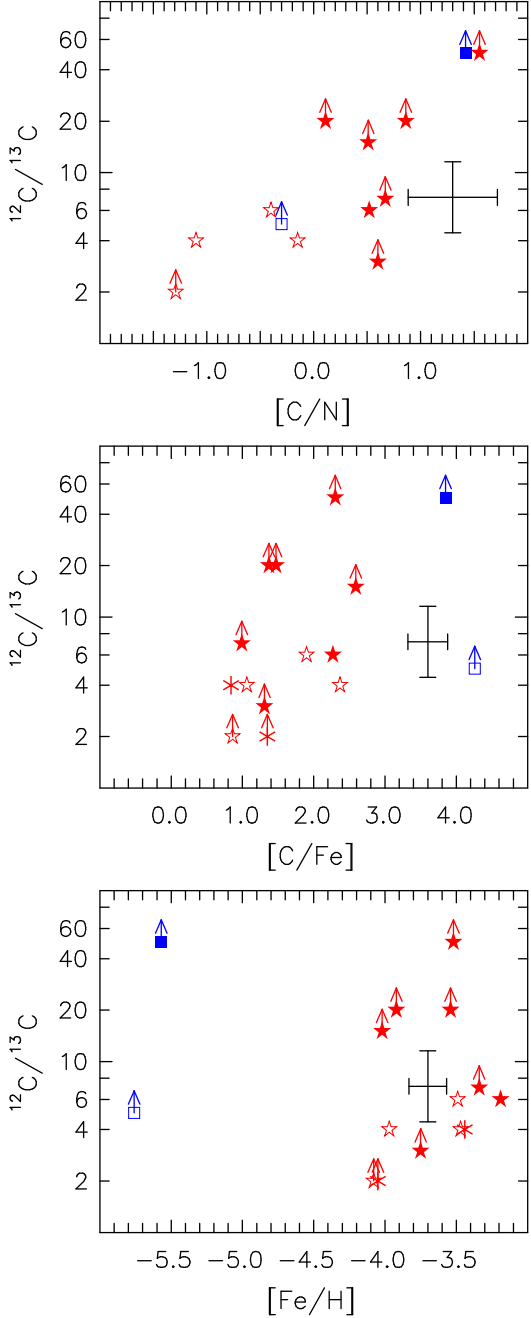}
  
\caption{\label{fig:c1213} \small $^{12}$C/$^{13}$C (logarithmic)
  vs. [C/N], [C/Fe], and [Fe/H] for C-rich stars.  The symbols are as
  defined in Figure~\ref{fig:cno}.}

\end{center}
\end{figure}

\subsubsection{{\hez} -  an NEMP object?}

In the context of the creation of CEMP stars via mass transfer from an
AGB companion, \citet{johnson07} noted that very metal-poor
intermediate-mass AGB stars are expected to produce large amounts of
primary nitrogen. They defined nitrogen-enhanced metal-poor (NEMP)
stars to be those with +0.5 $<$ [C/Fe] $<$ +1.0, [N/Fe] $>$ +0.5 and
[C/N] $<$ $-$0.5, and sought to find such stars.  In their sample of
21 objects, they expected to find 12\% -- 35\% to be NEMP stars. None,
however, was found, and only four stars in the recent literature
could be classified as possible NEMP stars.  One of their four
candidate NEMP stars was CS~22949-037, a CEMP-no star in our
Table~\ref{tab:basic_cempno}, having [N/Fe] = +2.2 and [C/N] = --0.9.
As discussed by \citet{johnson07}, the O abundance in this object is
higher than expected relative to the N abundance, assuming that the N
and O come from a companion AGB star. Combined with the lack of
s-process enrichment, they suggested that this star is not the result
of AGB pollution.

There is a second star in Table~\ref{tab:basic_cempno}, the red giant
{\hez}, with {\teff} = 5260\,K, {\logg} = 2.6, and [Fe/H] =--4.1,
which has [C/Fe] = +0.9, [N/Fe] = +2.2 and [C/N] = $-$1.3, and thus
also satisfies the above NEMP criteria.  Although we do not measure
the O abundance in {\hez}, the lack of s-process enrichment seems at
odds with the AGB-pollution scenario envisaged by
\citeauthor{johnson07} as producing NEMP stars.  That said, the
nucleosynthetic yields of the s-process elements by AGB stars at the
lowest metallicities remain uncertain due to limitations in the
modeling\footnote{One might add that this is not too surprising given
  the {\it ad hoc} introduction of the ``carbon pocket'' into models
  of AGB evolution in order to produce s-process enhancements at
  higher metallicities (--2.0 $<$ [Fe/H] $<$ --1.0).  See additional
  discussion of the modeling of the abundance patterns of CEMP-s stars
  in \citet[and references therein]{bisterzo12}.}.  An alternative
explanation of relative nitrogen richness in CEMP-no stars such as
CS~22949-037 and HE~0057--5959 may be afforded by stellar evolution
involving rapid rotation and ``mixing and fallback'' SN explosions,
which we shall discuss in Section~\ref{sec:discussion}.  We shall also
consider the question of the binarity of the C-rich stars in
Section~\ref{sec:binarity}.

We note for completeness that {\hez} also possesses an anomalously
high lithium abundance. We determine A(Li) =
$\log{\epsilon}{\rm{(Li)}} = \log(N_{\rm Li}/N_{\rm H}) + 12$ = 2.12.
Metal-poor red giants of similar {\teff} and {\logg} generally have
considerably lower values than this, A(Li) $\la1$, as the result of Li
destruction in their convective envelopes (see, e.g., \citealp[their
  Figure 5]{lind09}).  We shall return to this matter in Paper V
(Norris et al.\ 2013, in prep.).  The reader may be interested in the
fact that the other potential NEMP candidate, CS~22949-037, referred
to above (which is also a red giant, with {\teff} = 4960\,K and
{\logg} = 1.8) does not share this anomaly; lithium is not detected in
this star \citep{depagne02}.

\subsection{Relative Abundances [X/Fe] as a Function of [Fe/H], [C/Fe], and [Mg/Fe]}

\subsubsection{The enhancements of Na, Mg, Al, Si, and Ca}

\citet{aoki02b} first highlighted the large enhancements of Mg in
CEMP-no stars, reporting that CS~22949-037 and CS~29498-043 have
[Mg/Fe] = +1.38 and +1.52, respectively.  They also noted that [Al/Fe]
and [Si/Fe] are enhanced in both objects, while data from
\citet{mcwilliam95} and \citet{aoki04} show a similar effect for
      [Na/Fe].  Other C-rich stars exhibit this phenomenon: (i) the
      C-rich, most Fe-poor star, {\hea} shows extreme enhancements
      relative to Fe for these elements, with [Na/Fe] = +2.48, [Mg/Fe]
      = +1.55, and [Al/Fe] = +1.23 (no abundance estimate is available
      for Si), and (ii) \citet{cohen08} have commented on the Na, Mg,
      and Al enhancements in the C-rich stars HE~2323--0256
      (CS~22949-037) and HE~1012--1540 in Table~\ref{tab:abundances}.

Figure~\ref{fig:mgc} presents the dependence of [Na/Mg], [Mg/Fe],
[Al/Fe], [Si/Fe], and [Ca/Fe], as a function of [Fe/H] and [C/Fe]
(left and middle columns), based on the data in
Table~\ref{tab:abundances}.  The right column presents the generalized
histograms of [X/Fe] for the C-rich stars.  The outstanding and
remarkable feature of the figure is that all of Na, Mg, Al, and Si are
enhanced in approximately half of these stars, while Ca is enhanced in
only four of the 18 (22\%) for which data are
available.\footnote{These estimates are based on the somewhat
  subjective assessment that a star has an enhancement of element X if
  its LTE, 1D abundances satisfy [X/Fe] $>$ +0.8 (Na), +0.8 (Mg),
  --0.2 (Al), +0.8 (Si), and +0.6 (Ca).}  In Figure~\ref{fig:mgc}, one
also sees that the range in the abundance spreads decreases with
atomic number as one progresses from Na to Ca.  We shall return to
this point in the following section.  There are also strong
correlations between the enhancements of Na, Mg, Al, and Si, and also
of O, as is shown in Figure ~\ref{fig:omg}, where relative abundances,
[X/Fe], of O, Na, Al, Si, and Ca are presented as a function of
[Mg/Fe].

\begin{figure*}[!tbp]
\begin{center}
\includegraphics[width=14.0cm,angle=0]{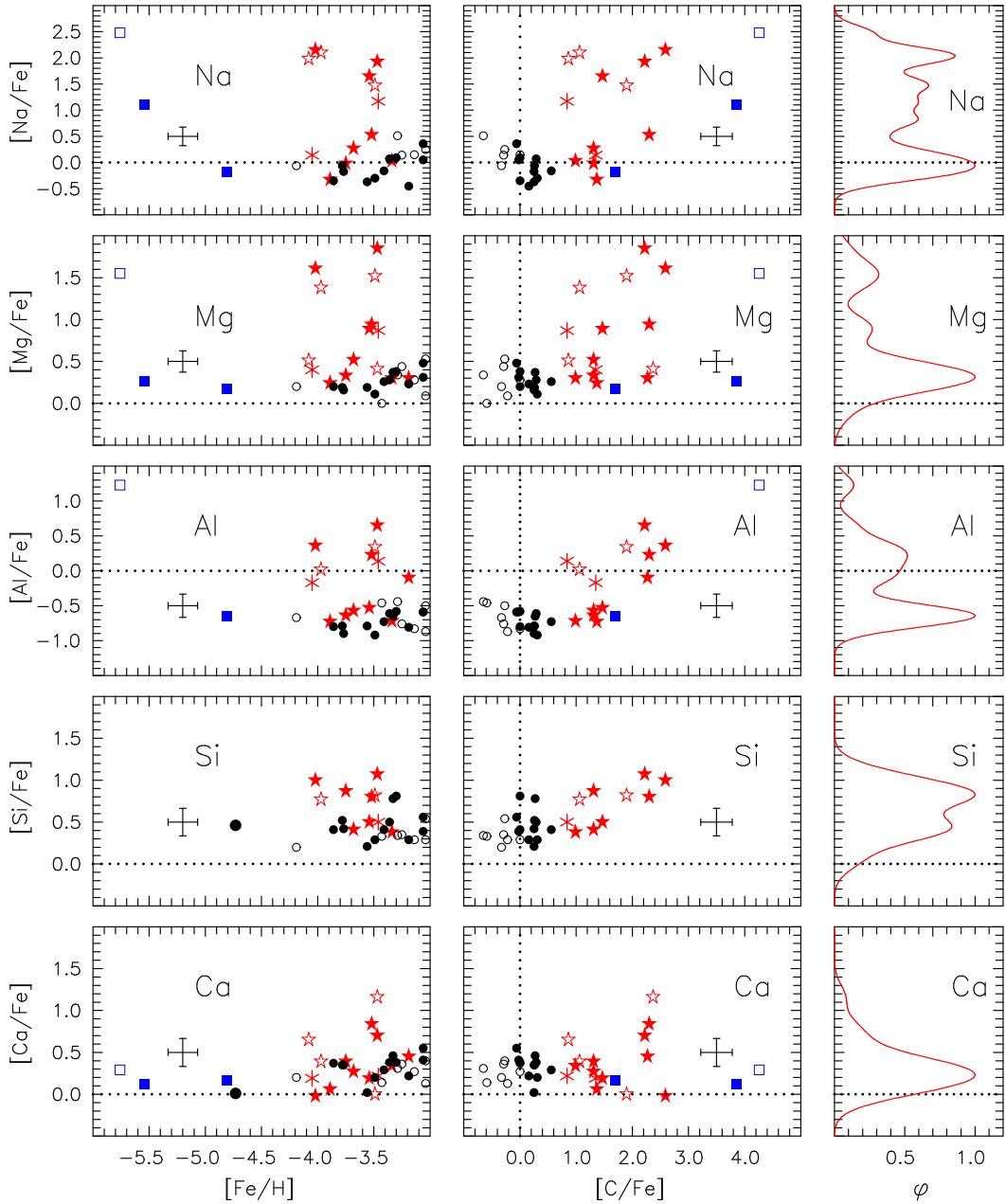}

\caption{\label{fig:mgc} \small Relative abundances of Na, Mg, Al, Si,
  and Ca vs. [Fe/H] (left column) and [C/Fe] (middle column) for
  C-rich and C-normal stars having [Fe/H] $<$ --3.0. The symbols are
  as defined in Figure~\ref{fig:cno}.  The right-hand panels present
  generalized histograms of [X/Fe] for the C-rich stars, obtained by
  using a gaussian kernel having $\sigma$ = 0.15 dex (the histograms
  have been normalized to unity). See text for discussion.}

\end{center}
\end{figure*}

\begin{figure}[!tbp]
\begin{center}
\includegraphics[width=6.7cm,angle=0]{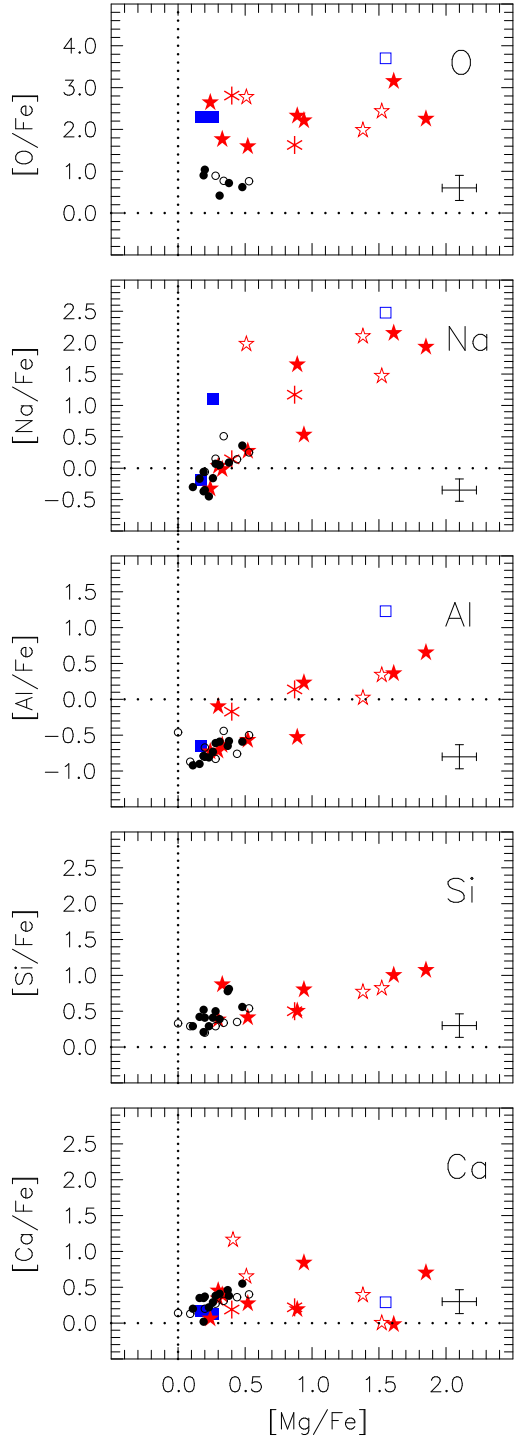}
  
\caption{\label{fig:omg} \small Abundances of O, Na, Mg, Al, Si, and
  Ca, relative to Fe, as a function [Mg/Fe], for C-rich and C-normal
  stars having [Fe/H] $<$ --3.0. The symbols are as defined in
  Figure~\ref{fig:cno}.  Note the strong correlations, diminishing as
  atomic number increases.  See text for discussion. }

\end{center}
\end{figure}

These are fundamental results to which we shall return in
Section~\ref{sec:discussion}.  

\subsection{Abundances as a Function of Atomic Number}

Figures~\ref{fig:xfez1} and \ref{fig:xfez2} provide alternative
representations of the data, and show the dependence of relative
abundance, [X/Fe], on atomic number, Z, for 20 of the C-rich stars in
Table~\ref{tab:basic_cempno}.  (We exclude three stars in the table
that have [Fe/H] $>$ --3.0.)  (In these figures the abundance errors
are commensurate with the size of the symbols.)  For comparison
purposes, the line in each panel shows data for a C-normal star having
the same {\teff}/{\logg}/[Fe/H], following \citet{yong12}\footnote{For
  the three stars with [Fe/H] $<$ --4.5, the reference C-normal stars
  have [Fe/H] = --4.2.}. There are two points worth making.  First,
there is little evidence for non-solar relative abundances ([X/Fe])
for elements with 20 $\la$ Z $\la$ 28; almost all of the large
variations occur for Z $<$ 20.  Second, the enhancements become
larger, on average, as [Fe/H] decreases: the largest variations occur
below [Fe/H] = --3.4.  We highlight this by circling the Mg values in
the two figures, and note that while eight of the 16 stars with [Fe/H]
$<$ --3.4 have [Mg/Fe] $>+0.8$, none of the seven stars with [Fe/H]
$>$ --3.4 in Table~\ref{tab:basic_cempno} (of which five are presented
in Figures~\ref{fig:xfez1} and \ref{fig:xfez2}) has [Mg/Fe] $>$ +0.4.
(We also note that examination of the abundances of the CEMP-no stars
of \citet{barklem05} presented in Table~\ref{tab:barklem}, all of
which have [Fe/H] $>$ --3.5, shows that none of them has [Mg/Fe] $>$
+0.6.)  We conclude that the relative abundance [Mg/Fe] becomes
larger, on average, as [Fe/H] decreases.

\begin{figure*}[!tbp]
\begin{center}
\includegraphics[width=14.0cm,angle=0]{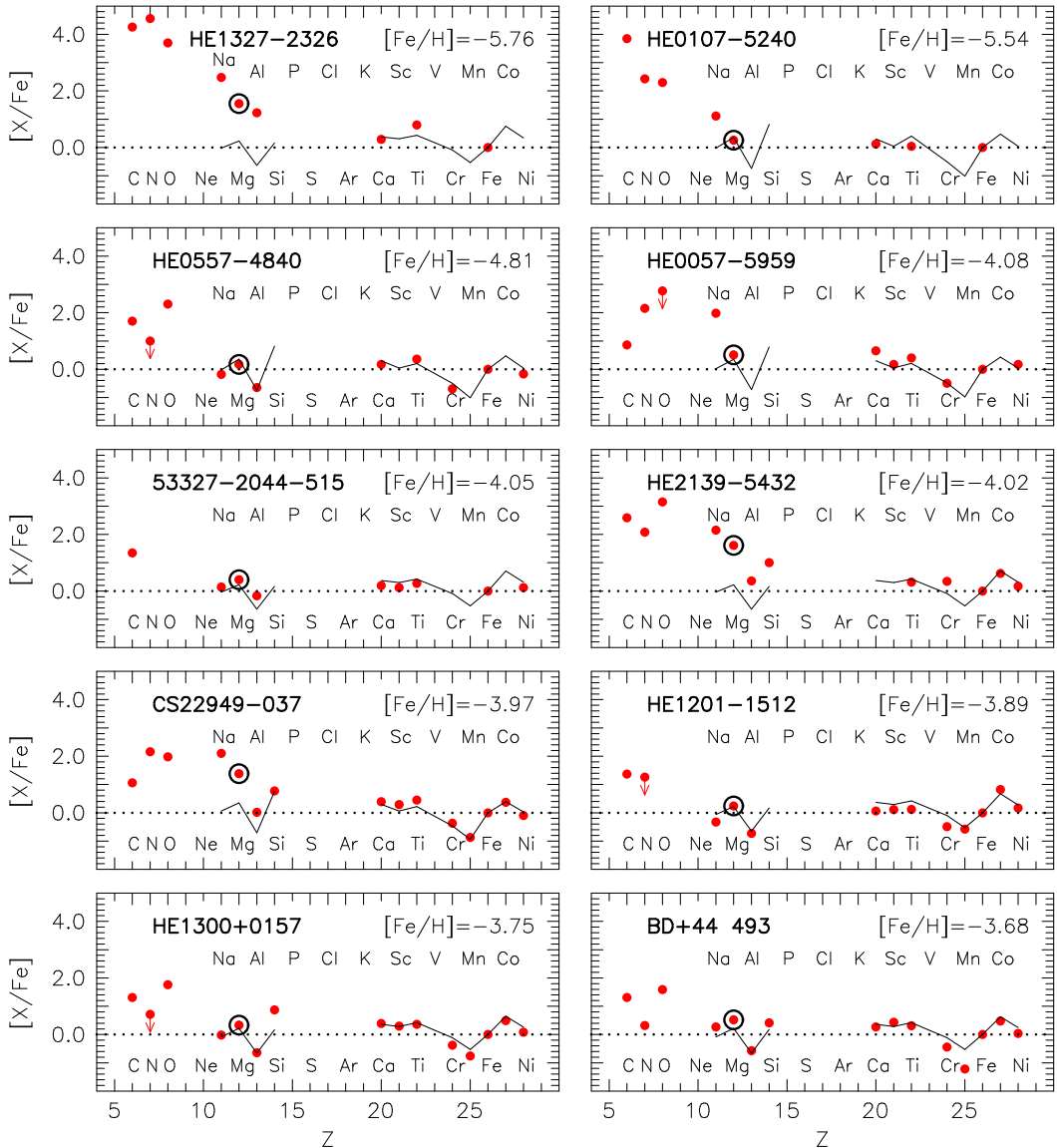}
 
\caption{\label{fig:xfez1} \small Relative abundances, [X/Fe],
  vs. atomic number, Z, for the C-rich stars in
  Table~\ref{tab:abundances} having [Fe/H] $<$ --3.7.  The most
  Fe-poor stars are presented in the top panels; [Fe/H] increases from
  top to bottom.  The line in each panel represents data for a
  C-normal star having the same {\teff}/{\logg}/[Fe/H] values as the
  C-rich star. Note the enormous overabundances of the relative
  abundances of the light elements, decreasing to solar values (the
  dotted horizontal lines) for Z $>$ 20.}

\end{center}
\end{figure*}

\begin{figure*}[!tbp]
\begin{center}
\includegraphics[width=14.0cm,angle=0]{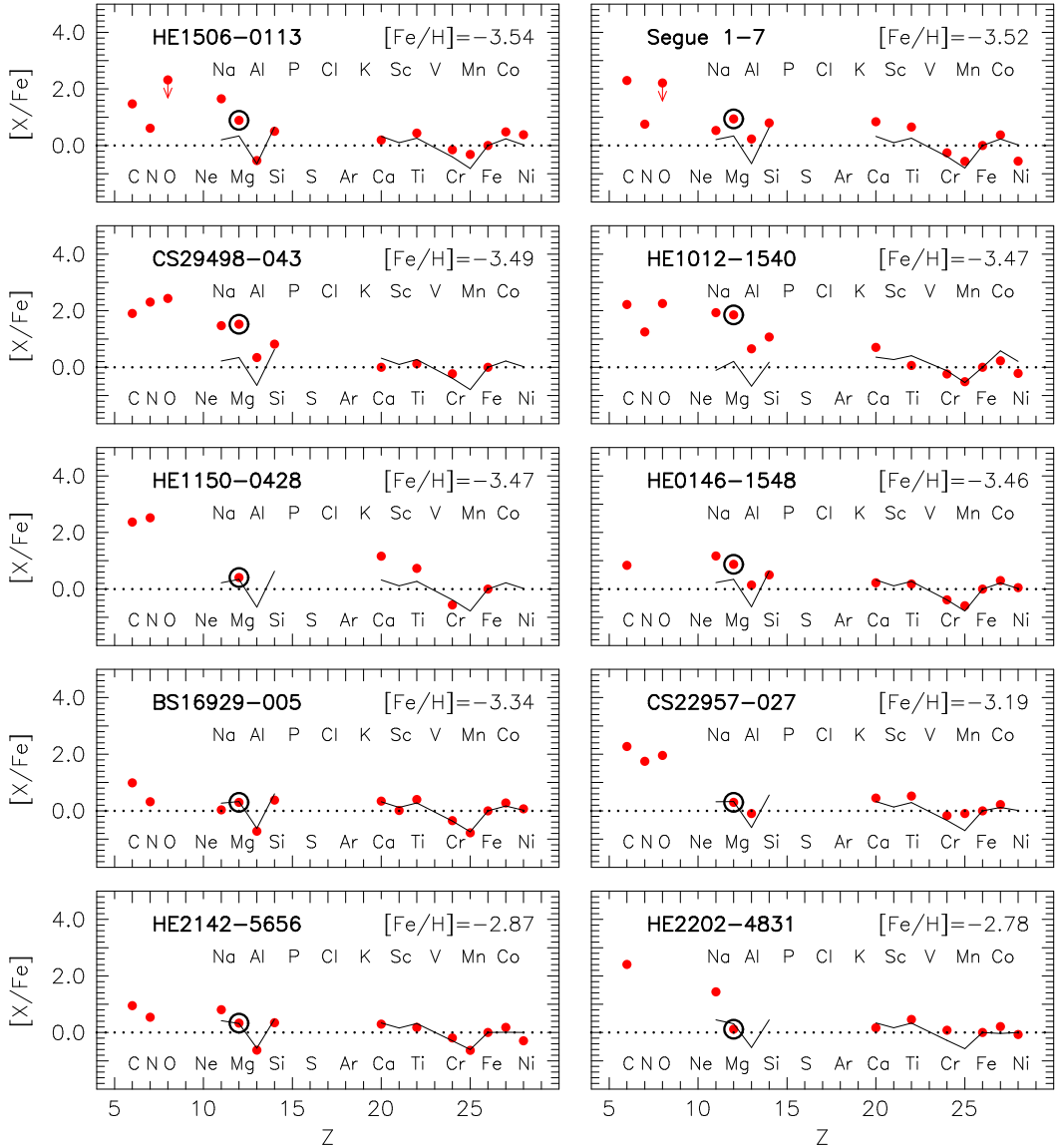}
  
\caption{\label{fig:xfez2} \small The same as for
  Figure~\ref{fig:xfez1}, for C-rich stars in
  Table~\ref{tab:abundances} having [Fe/H] $>$ --3.6. }

\end{center}
\end{figure*}

We pursue the dependence of abundance enhancement, as a function of
[Fe/H], in Figure~\ref{fig:dxfehis}, and for elements Na -- Ba,
present generalized histograms of $\Delta$[X/Fe] = [X/Fe]$_{\rm
  C-rich}$ -- [X/Fe]$_{\rm C-normal}$ and $\Delta$[Sr/Ba] =
[Sr/Ba]$_{\rm C-rich}$ -- [Sr/Ba]$_{\rm C-normal}$, the enhancement of
the ratio in C-rich stars above the values of C-normal stars having
the same atmospheric parameters, {\teff}/{\logg}/[Fe/H].  In each
panel the full line represents stars having [Fe/H] $\le-$3.4, while
the dotted one is for stars with [Fe/H] $>$ --3.4.  Here one sees
differences between the Fe-poorer and Fe-richer histograms in each
panel that decrease as one progresses from Na to Ca.  For Sc through
Ni there is little evidence for differences.  We have included results
in the figure for Sr and Ba for completeness.  One sees large spreads
for Sr for both Fe groups, but only relatively smaller ones for Ba.
Given the large spreads that exist for these two elements in C-normal
stars, in particular for Sr (see, e.g., Figures 29 and 30 of Paper
II), one might wonder about our attempt to define an excess of these
elements in C-rich stars relative to values in C-normal objects.  We
shall not consider the matter further here.

\begin{figure*}[!tbp]
\begin{center}
\includegraphics[width=10.0cm,angle=0]{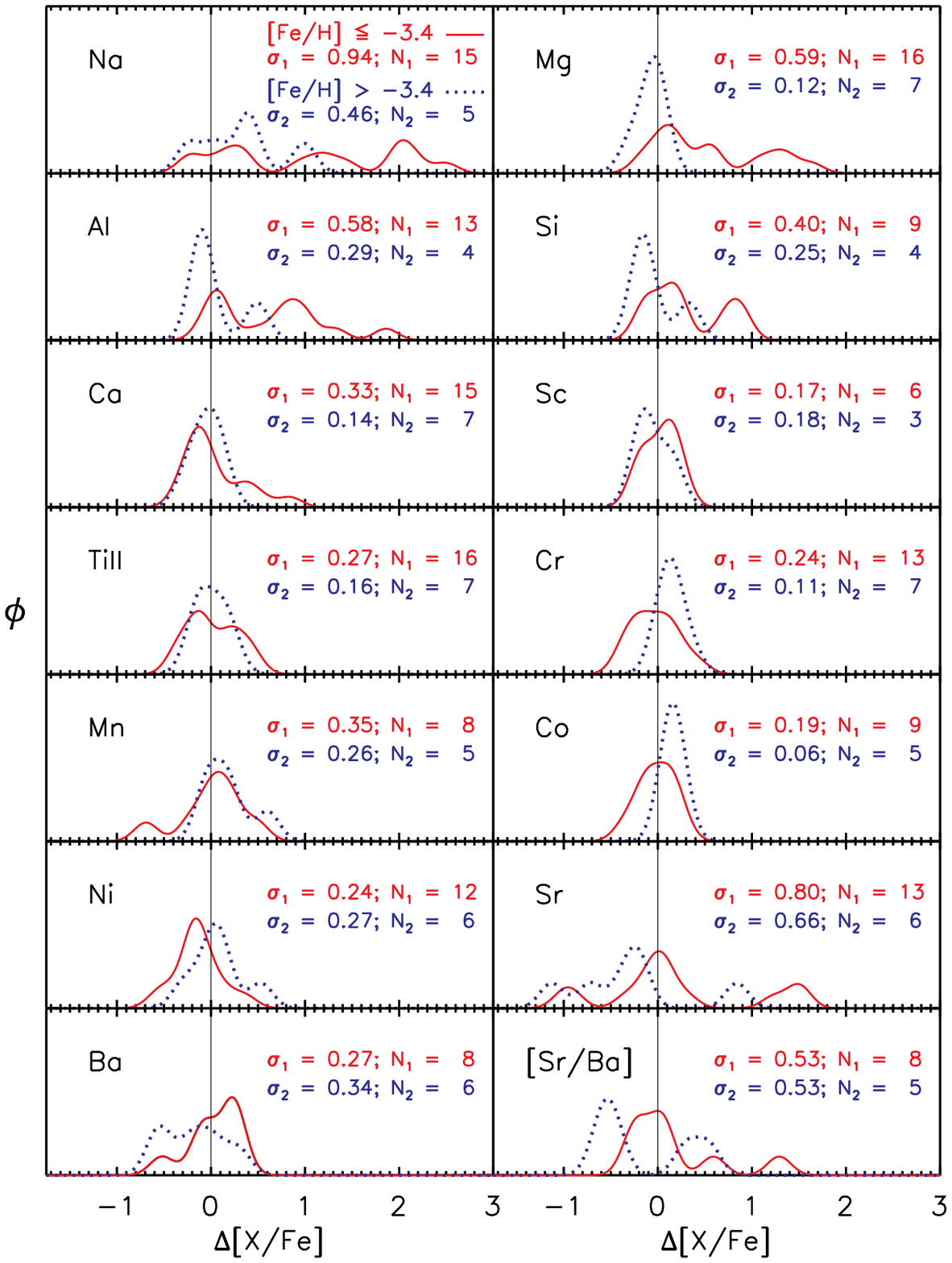}
  
\caption{\label{fig:dxfehis} \small Generalized histograms of
  $\Delta$[X/Fe] = [X/Fe]$_{\rm C-rich}$ -- [X/Fe]$_{\rm C-normal}$
  and $\Delta$[Sr/Ba] = [Sr/Ba]$_{\rm C-rich}$ -- [Sr/Ba]$_{\rm
    C-normal}$, the enhancement of the ratios in C-rich stars above the
  values of C-normal stars having the same atmospheric parameters
  {\teff}/{\logg}/[Fe/H], obtained using a gaussian kernel having
  $\sigma$ = 0.30 dex.  In each panel the full line represents stars
  having [Fe/H] $\le-$3.4, while the dotted one is for stars with
  [Fe/H] $>$ --3.4. (The scalings have been chosen so that the area
  enclosed by all histograms is the same.)}

\end{center}
\end{figure*}

As noted in the previous section, the degree of enhancement of Na
through Ca in C-rich stars appears to decrease with increasing atomic
number.  This is confirmed in Figure~\ref{fig:dxfehis}.  Specifically,
for [Fe/H] $\le-$3.4, the dispersions of $\Delta$[X/Fe] for Na, Mg,
Al, Si and Ca in the C-rich stars are 0.94 $\pm$ 0.17 dex, 0.59 $\pm$
0.10 dex, 0.58 $\pm$ 0.11 dex, 0.40 $\pm$ 0.09 dex, and 0.33 $\pm$
0.06 dex, respectively.

It should be emphasized that not all C-rich stars exhibit enhancements
of Na, Mg, Al, and Si.  We shall return to this in
Section~\ref{sec:discussion}, and argue that the spreads in [X/Fe]
observed for these elements may be a natural result of two of the
explanations for the C-rich stars proposed in the literature.

\section{THE INCIDENCE OF BINARITY AMONG THE C-RICH STARS WITH [FE/H] $\la-3.0$ }\label{sec:binarity}

Binarity has been suggested as a necessary or likely explanation for
the carbon richness of some or all of the C-rich stars (21 CEMP-no
stars plus two having [Fe/H] $\sim$ --5.5) in
Table~\ref{tab:basic_cempno} (e.g., \citealp{suda04};
\citealp{masseron10}).  To our knowledge, only one CEMP-no star,
CS~22957-027 \citep{preston01}, is known to exhibit radial velocity
variations.  Of the other stars in our Table~\ref{tab:basic_cempno},
we are aware of detailed observations of only {\ito}, which was
extensively monitored for variations by \citet{carney03}.  They
reported a velocity dispersion of $\sigma$ = 0.8 {\kms} and a velocity
range of 3.2 {\kms} from 28 observations spanning 4982 days, and did
not classify it as binary.  That said, given the long histories needed
to establish the universality of binarity among the Ba and CH stars
\citep{mcclure90} and the CEMP-s stars \citep{lucatello05b}, one
should be very hesitant to rush to judgment on the issue of
variability for the C-rich class under discussion here.

In Table~\ref{tab:radvel}, we summarize the results of our literature
search for radial velocity measurements of the 23 C-rich stars in
Table~\ref{tab:basic_cempno}.  Columns (1) -- (3) contain star name,
average heliocentric radial velocity, and number of
epochs\footnote{Given the long periods and timespans involved, we
  average velocities, if taken within an interval of $\sim2$~days, to
  obtain an individual velocity observed at the ``epoch'' defined by
  the average of the individual times of observation.} for which
velocity data are available, while columns (4) -- (6) present the
observed velocity range for each object, the span (in days) of the
observations, and the data sources, respectively.  (In the final
column of the table we present distances for those stars having [Fe/H]
$<$ --3.1, which we shall introduce and use in
Section~\ref{sec:kinematics}.)  Taken at face value, among the 13
stars in Table~\ref{tab:radvel} with multiple observations,
CS~22957-027 is the only one ($\sim$8\% of the sample) for which
variations greater than $\sim3$~{\kms} have been observed.  For
comparison, we note that \citet{carney03} report the spectroscopic
binary frequency for giants with [Fe/H]$\le -1.4$ and periods less
than 6000~days is 16 $\pm$ 4\%, and 17 $\pm$ 2\% for dwarfs of similar
metallicity.

\input{tab5}

What is the probability of observing the preponderance of the small
velocity ranges seen in Table~\ref{tab:radvel}, given the observed
numbers of epochs and their time spans?  We have addressed this issue
using Monte Carlo simulations, as follows.  We first assumed that each
star has an observed sinusoidal radial velocity curve with
semi-amplitude 10~{\kms} and period 3125~days, similar in first
approximation to the values observed for CS~22957-027 by
\citet{preston01}.  We excluded CS~22957-027 and {\ito}, and for each
of the other 11 stars in the table with multiple observations set the
first ``observation'' at a random phase for which we determined the
velocity, and then obtained the velocities that would be observed at
all other epochs of observation of that star.  We also assumed that
the orbital plane of the binary was inclined at random with respect to
the plane of the sky, and determined the individual velocity range
expected for each of the 11 stars.  We repeated the exercise 100,000
times and asked ``In what fraction of the exercises would the
simulated observations of each star exhibit a velocity range no
greater than that actually observed or 1.0 {\kms}, whichever was the
larger''\footnote{We set a lower limit 1.0 {\kms}, which corresponds
  to the error in the difference of the two velocities that determine
  the velocity range in column (4) of Table~\ref{tab:radvel}, each of
  which is assumed to have an error of measurement of 0.7
  {\kms}.}. The fraction was 0.0.  No case was obtained in which all
11 stars exhibit radial velocity variations smaller than the larger of
the observed range and 1.0 {\kms}.  We then asked the question ``At a
given assumed velocity curve semi-amplitude of the putative binary,
what is the period greater than which the above Monte Carlo process
would `observe' no velocity span, for each star, greater than the
larger of the span presented in column (4) of Table~\ref{tab:radvel}
and 1.0~{\kms}, in 99\% of cases''.  The results of this exercise are
presented in Figure~\ref{fig:plap}, where the continuous line
represents the locus of the velocity curve semi-amplitude, K$_{1}$, as
a function of the resulting value of log(Period).  For comparison, we
also include data for the observed positions of CH stars (triangles;
\citealp{mcclure90}), CEMP-s stars (filled circles;
\citealp{lucatello05b}), and the binary CEMP-no star, CS~22957-027
(star symbol), from Table~\ref{tab:radvel}.  For the interest of the
reader we also plot the position of the \citet{suda04} binary model
for the C-rich star {\hen} (period 150 yrs, K$_{1}$ = 7 {\kms}).  To
the left of the line, the observed velocity ranges in
Table~\ref{tab:radvel} exclude, at the 99\% level, the hypothesis that
all of the C-rich stars with [Fe/H] $\la-3.0$ are binary, while to the
right the hypothesis is accepted.  The periods that are consistent
with the null hypothesis are very long: P $\ga$ 10,000 days = 27 yrs.
It may be some time before this issue is settled.

\begin{figure}[!tbp]
\begin{center}
\includegraphics[width=7.5cm,angle=0]{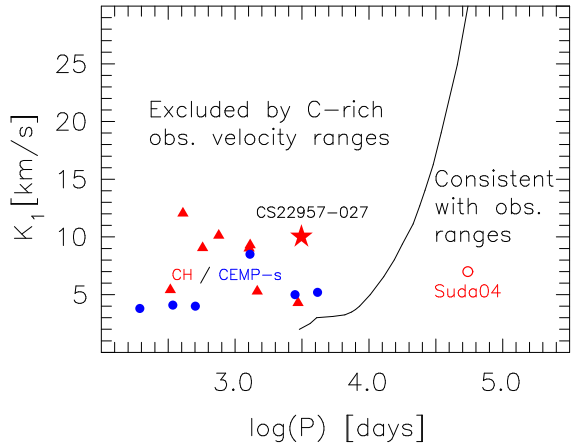}
  
\caption{\label{fig:plap} \small Radial-velocity curve semi-amplitude
  K$_{1}$ vs log(Period) for binary stars.  The continuous line
  separates regions where Monte-Carlo simulations reject (to the left)
  and accept (to the right) the null hypothesis that the small
  observed velocity ranges for C-rich stars (CEMP-no and two hyper
  metal-poor stars) of Table~\ref{tab:radvel} result from binarity.
  See text for discussion.  Observed values are presented for binary
  CH-stars \citep{mcclure90} (triangles) and CEMP-s stars
  \citep{lucatello05b} (filled circles), while the CEMP-no star
  CS~22957-027 \citep{preston01} is plotted as a star.  The open
  circle shows the \citet{suda04} model values for the most Fe-poor
  red giant {\hen}.}

\end{center}
\end{figure}

In summary, the binary statistics for CEMP-no stars are decidedly
different from those of CEMP-s stars.  Further data are necessary to
more fully characterize the binary nature of the CEMP-no subclass.

\section{THE KINEMATICS OF THE C-RICH AND C-NORMAL STARS WITH [FE/H] $<-3.1$}\label{sec:kinematics}

The radial velocities presented in Table~\ref{tab:radvel} contain
information on the kinematics of the C-rich population in
Table~\ref{tab:basic_cempno}.  In order to investigate whether the
C-rich and C-normal populations have the same kinematic properties,
Table~\ref{tab:radvel_cnormal} presents radial velocities from the
literature for the 34 C-normal stars in Paper II, plus {\caffau11},
that have [Fe/H] $<$ --3.1.  As before, we define C-normal as [C/Fe]
$\le+0.7$.  The contents of columns (1) -- (5) have been taken from
Paper II, while columns (6) -- (7) contain the radial velocities
together with their sources. We would agree with the critic who
suggests that our choice of the upper limit of [Fe/H] = --3.1 is
somewhat arbitrary; that said, we would note in reply that this value
concentrates our investigation on the regime where the C-rich
population of Table~\ref{tab:basic_cempno} is best defined, with
minimal contamination from the CEMP-r, -r/s, and -s subclasses. In the
final column of the table (and that of Table~\ref{tab:radvel}, as
foreshadowed in Section~\ref{sec:binarity}), we present distances
required in the kinematic analysis that follows.  These were obtained
by first fitting the {\teff}, {\logg}, and [Fe/H] values in
Tables~\ref{tab:basic_cempno} and ~\ref{tab:radvel_cnormal} to the
Yale--Yonsei Isochrones
\citep{demarque04}\footnote{http://www.astro.yale.edu/demarque/yyiso.html},
for an age of 12 Gyr, to obtain absolute magnitudes, M$_{\rm V}$, and
then using these in conjunction with apparent $V$ magnitudes and
E($B-V$) reddenings taken from the literature\footnote{We note two
  exceptions.  For the stars {\caffau} and Segue1-7 we adopt the
  distances of \citet{caffau11} and \citet{martin08}, respectively.}.

\input{tab6}

Following \citet{frenk80}, we determine, for these C-rich and C-normal
populations, the Galactic systemic rotational velocity, V$_{\rm rot}$,
and $\sigma_{\rm los}$, the rms value for the dispersion of the
line-of-sight peculiar motions with respect to the group motion.  We
refer the reader to \citet{norris86} and \citet{beers95} for our
earlier applications of this technique in the determination of the
systemic rotation of the older and more metal-poor populations of the
Galaxy.  For the 18 C-rich stars in Table~~\ref{tab:radvel} with
[Fe/H] $<$ --3.1 we obtain V$_{\rm rot}$ = --44 $\pm$ 45~{\kms} and
$\sigma_{\rm los}$ = 90 $\pm$ 15~{\kms}, while for the 35 C-normal
stars in Table~\ref{tab:radvel_cnormal} the corresponding numbers are
--76 $\pm$ 59~{\kms} and 151 $\pm$ 18~{\kms}, respectively.  The
differences between the two populations are $\Delta$V$_{\rm rot}$ = 32
$\pm$ 74~{\kms} and $\Delta\sigma_{\rm los}$ = 61 $\pm$ 23~{\kms}.
That is, the present results show no significant difference between
V$_{\rm rot}$ for the two groups, and a 2.6$\sigma$ difference between
their $\sigma_{\rm los}$ values.  While the latter result may be
considered suggestive of a real difference, we suggest that more data
should be obtained to test its reality.

\citet{carollo12} have demonstrated (i) the existence of a significant
increase in the fraction of CEMP stars with increasing height above
the Galactic plane, $|$Z$|$, and (ii) that the frequency of CEMP stars
associated with the outer-halo population is significantly higher than
that of the inner-halo.  We recall that their result was based on
material principally more metal-rich than [Fe/H] = --3.0: the most
metal-poor bin in their Figure 15 at [Fe/H] $\sim$ --2.7 contained
C-rich fractions of 20\% and 30\% for their inner- and outer-halo
components, respectively.  We recall also that \citet{carollo12},
given the spectral resolution of their spectra, were unable to
determine the CEMP subclass of stars in their data set.  A possible
explanation of their results is that it arises from a relatively
larger fraction of CEMP-no stars in the outer halo, in particular for
[Fe/H] $\la-2.0$.  A prediction of this conjecture is that the
CEMP-no/CEMP-s ratio was higher in the Galaxy's accreted dwarf
galaxies that preferentially populated its outer rather than its inner
regions.  Future determination of this ratio in the Galaxy's satellite
dwarf galaxies should be undertaken to constrain this possibility.

Application of the \citet{frenk80} formalism to the combined sample of
the 53 C-rich and C-normal stars having [Fe/H] $<$ --3.1, yields
V$_{\rm rot}$ = --64 $\pm$ 41~{\kms}, and $\sigma_{\rm los}$ = 133
$\pm$ 13~{\kms}.  We also divided this sample into two groups having
essentially equal size, the first containing the 26 stars with [Fe/H]
$\ge-3.5$ and the other the 27 with [Fe/H] $<$ --3.5.  The results are
presented in Table~\ref{tab:kinematics}, and plotted in
Figure~\ref{fig:kinematics} (star symbols) as a function of mean iron
abundance, $\langle$[Fe/H]$\rangle$.  For comparison purposes we also
include the results for the metal-poor halo samples of \citet[Table
  9]{norris86} (filled circles) and \citet[Table 3]{beers95} (open
circles).  The most interesting feature of the figure is the large
retrograde V$_{\rm rot}$ = --119 $\pm$ 64~{\kms} value at [Fe/H] =
--4.0, some 2$\sigma$ below the rotational velocity of $\sim$ 20
{\kms} of halo material in the range --1.5 $<$ [Fe/H] $<$ --3.0.  Once
again, further data are required before one can regard the result as
definitive.  Another reality check is provided by consideration of the
sample sizes used.  We refer the reader to \citet[Section
  III(d)]{norris86}, who addressed the issue by using simulated
samples, and concluded ``The data suggest that reliable results can be
obtained with all sample sizes, in the sense that the errors
accurately reflect the quality of the estimates but that if relatively
accurate information is required samples of size in excess of 100
objects are necessary''.

\begin{figure}[!tbp]
\begin{center}
\includegraphics[width=8.5cm,angle=0]{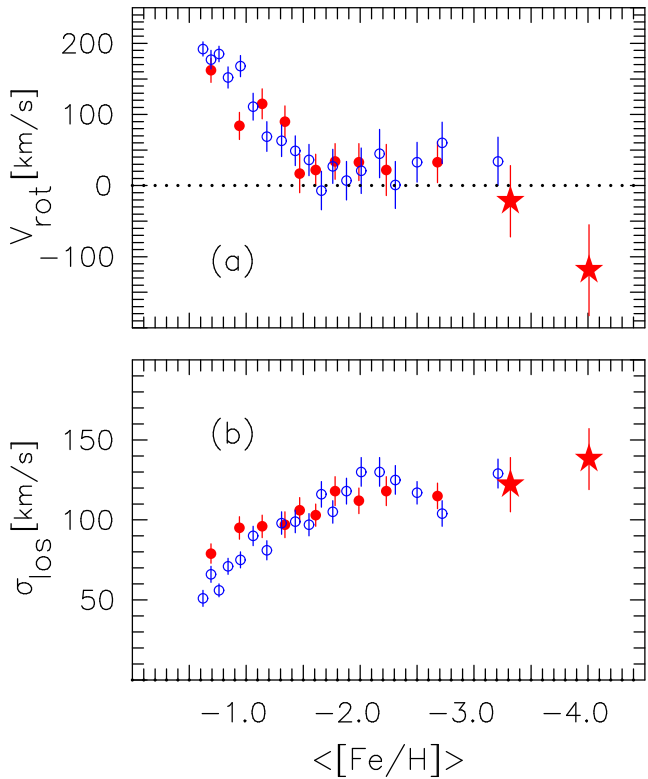}
  
\caption{\label{fig:kinematics} \small The dependence of (a) the
  systemic Galactic rotational velocity, V$_{\rm rot}$, and (b) the
  line-of-sight velocity dispersion, $\sigma$$_{\rm los}$, on [Fe/H],
  for the stars of the present work having [Fe/H] $<$ --3.1 (filled
  star symbols; from Table~\ref{tab:kinematics}), together with results
  for the metal-poor halo samples of \citet[Table 9]{norris86} (red
  filled circles) and \citet[Table 3]{beers95} (blue open circles).
  See text for discussion.}

\end{center}
\end{figure}

\input{tab7}

\section{DISCUSSION}\label{sec:discussion}

\subsection{Suggested Theoretical Scenarios} 

To this point, we have discussed various observational aspects of the
C-rich stars -- their chemical abundance patterns, binarity, and
kinematics.  A great amount of effort in the literature has also been
devoted, from a theoretical perspective, to understanding the origin
of the most metal-poor stars, and in particular to the large fraction
that is C-rich.  We now turn our attention to these efforts, in an
attempt to understand the possible origin(s) of the C-rich stars, and
briefly outline some of the suggestions that have been made.  This
topic has been considered at some length, and we refer the reader to
the works of \citet{beers05}, \citet{cohen08}, \citet{frebel07},
\citet{fujimoto00}, \citet{heger10}, \citet{joggerst10},
\citet{lai08}, \citet{masseron10}, \citet{meynet06, meynet10},
\citet{nomoto06}, \citet{norris07}, and \citet{suda04} for previous
discussions.  Here are some of the phenomena suggested to be involved
in various chemical enrichment scenarios of material initially having
zero or very low heavy-element content.

\begin{enumerate}

\item {\it Fine-structure line transitions of C\,II and O\,I as a
  major cooling agent in the early Universe (\citealp{bromm03})} 

C- and/or O- rich material forms stars, through fragmentation, on shorter
timescales than in regions where their abundances were lower, leading
to the formation of long-lived, low mass C- and/or O-rich stars still
observable today.

\item {\it Supermassive (M $>$ 100 M$_\odot$), rotating stars
  (\citealp{fryer01})}

In some mass ranges, rotation leads to mixing, by meridional
circulation, of C and O from the He-burning core into the H-burning
shell, leading to large N enhancements.

\item {\it Multiple generations of Type II supernovae (SNe) involving
  ``fallback'' (M $\sim10-40$ M$_\odot$) (\citealp{limongi03})}

The ejecta from a ``normal'' SN is combined with that from one of low
energy in which the outer layers (rich in light elements) are expelled,
while much of the inner layers (rich in the heavier elements) ``fall
back'' onto the central remnant.

\item {\it ``Mixing and fallback'' Type II SNe (M $\sim10-40$ M$_\odot$)
  (\citealp{umeda03,umeda05})}

Low energy SNe eject material preferentially from their outer regions,
which are enhanced in light elements, with the expulsion of only
relatively small amounts of the heavier elements formed deeper in the
star.  During the explosion, internal mixing occurs in an annulus
outside the mass cut at which the expansion in initiated.  A small
amount of mixed material is eventually expelled from the star, with
most of it falling back into the central regions.

\item {\it Type II SNe with relativistic jets (\citealp{tominaga07b})}

A relativistic jet-induced black-hole-forming explosion of a
40~M$_\odot$ SN leads to infall of inner material that ``decreases the
[ejected] amount of inner core material (Fe) relative to that of outer
material (C)''.

\item {\it Zero-metallicity, rotating, massive ($\sim60$~M$_\odot$)
  and intermediate mass ($\sim7$~M$_\odot$) stars (\citealp{meynet06,
    meynet10, hirschi07})}

Rotationally-driven meridional circulation leads to CNO enhancements
and large excesses of $^{13}$C (and hence low $^{12}$C/$^{13}$C
values), Na, Mg, and Al, in material expelled in stellar winds.  The
essential feature of rotation is to admix and further process the
products of H and He burning.

For investigations of the combined effects of mixing, fallback, and
rotation in massive stars over wide parameter ranges, we refer the
reader to \citet{heger10} and \citet{joggerst10}.

\item {\it Nucleosynthesis and mixing within low-mass, low-metallicity, stars
  (\citealp{fujimoto00, campbell10})}

Carbon is mixed to the outer layers of low-mass, extremely metal-poor
giant stars, while mixing -- driven by a helium flash -- transports
protons into the hot convective core.  Enhancements of Na, Mg, Al, and
heavy neutron-capture elements are also predicted.

\item {\it Population III binary evolution with mass transfer, and
  subsequent accretion from the interstellar medium (\citealp{suda04,
    campbell10})}

The primary of a zero-heavy-element binary system is postulated to
transfer C- and N-rich material, during its AGB phase, onto the
currently-observed secondary, which later accretes Fe from the
interstellar medium to become a CEMP star.

\item {\it Separation of gas and dust beyond the stellar surface
  during stellar evolution, followed by the accretion of the resulting
  dust-depleted gas (\citealp{venn08})}

The peculiar abundance patterns result from fractionation of the
elements onto grains, as determined by their condensation
temperatures, during stellar evolution, rather than being due to
``natal variations''.  Subsequent examination of the critical elements
sulfur and zinc in the Fe-poor, C-rich stars CS~22949--037 ([Fe/H] =
--4.0) and {\hea} ([Fe/H] = --5.8) by \citet{spite11} and
\citet{bonifacio12}, respectively, shows that they are detected in
neither object.  In {\hea}, the limits on [S/H] and [Z/H] are
consistent with the condensation hypothesis.  For CS~22949--037,
however, the limits for both elements lie some 1.5~dex below the
values that would be expected.  Given these results, we shall not
consider this mechanism further.  That said, it would be very valuable
to obtain further sulphur and zinc abundances, or limits, of more
C-rich stars, to examine the question in greater detail.  

\end{enumerate}

\subsection{Comparison with Theoretical Predictions}

In the context of the material presented in
Sections~\ref{sec:abundances} -- \ref{sec:kinematics} we now ask:
which of these mechanisms do the observations require; which, if any,
may be rejected; and which need further work to enable sharper
confrontation between observation and theory?  We consider the
observational constraints set by the abundance patterns and kinematics
of the C-rich stars in Table~\ref{tab:basic_cempno}.

\begin{enumerate}

\item {\it The ubiquitous CNO enhancements and low $^{12}$C/$^{13}$C
  values, the Na, Mg, Al, Si enhancements in $\sim50$\% of the
  population, and the relative normality of the heavier elements (Z
  $>$ 20)}

While CNO enhancements and low $^{12}$C/$^{13}$C values may originate
in several environments, those of Na, Mg, and Al are best explained in
terms of the bringing together and processing of material from
H-burning and He-burning regions in the intermediate depths of massive
and/or intermediate mass stars.  This is a generic property of the
``mixing and fallback'' models; of the zero-heavy-element, rotating,
massive and intermediate-mass stars; and of the Type II SNe with
relativistic jets scenarios (discussed above), all of which lead to
the expulsion of large amounts of these elements from the intermediate
depths in stars where they are produced.  See, for example, the model
enhancements of Na, Mg, and Al produced by \citet[Figures 1 and
  2]{iwamoto05} (``mixing and fallback''), and by \citet [Figures 8
  and 10]{meynet06} (for zero-heavy-element, rotating, massive and
intermediate mass stars).  Currently (to our knowledge), a comparison
of observation with theory is not available for Si and Ca for the
fast-rotating models.

The fact that only half of the C-rich stars exhibit large Na, Mg, and
Al enhancements seems readily explainable.  In ``mixing and fallback''
models, it results from the admixing of different radial zones, their
nuclear burning, and the expulsion of material that contains different
relative amounts of synthesized Na, Mg and Al.  We refer the reader to
\citet{iwamoto05} for an explanation of the different abundance
patterns of {\hen} and {\hea}.  For the ``fast rotator'' hypothesis,
one might naturally expect the relative amounts of Na, Mg, and Al
(products of H- and He-burning) to be a function of rotational
velocity.

\item {\it Enhancements of Si and Ca exist in some stars, and are
  relatively small compared with those of C, N, O, Na, Mg, and Al}

As noted above, the chemical enrichment produced by nucleosynthesis in
zero-heavy-element models of ``mixing and fallback'' SNe, of rotating,
massive and intermediate mass stars, and of SNe with relativistic jets
best explains the relative abundances presented here.  There is,
however, a basic difference between the rotating star models, on the
one hand, and the ``mixing and fallback'' and relativistic jet models,
on the other.  In principle at least, the two cases sample different
regions of the stars that produce the enrichment.  In the rotating
models, the regions providing the enrichment are the outer layers that
mix via meridional circulation, and much of the ejecta are expelled in
stellar winds, before exhaustion of the nuclear fuel in the central
regions leads to a potential explosion.  In the other class of model,
all enrichment patterns are determined in the supernova phase, during
which there is mixing and expulsion, potentially at least, of material
from all parts of the star outside the core.  Insofar as Si and Ca are
produced deeper in a star than are the lighter elements, they present
the potential to test the predictions of the different models more
closely.  In particular, it would be interesting to have more accurate
abundances of these elements in a larger sample of C-rich stars for
comparison with more detailed predictions of the two classes of
models.  This could be a very useful avenue for investigation.

\item {\it The increasing fractions of stars enhanced in C, N, O, Na,
  Mg, Al, and Si relative to Fe as [Fe/H] decreases}

The processes leading to the C-rich class appear to dominate at the
lowest values of [Fe/H]. We suggest that below [Fe/H] $\sim$ --3.0 the
data are consistent with the existence, among the stars we observe
today, of two populations that have quite distinct abundance patterns
of the light elements, with the observed C-rich population being the
more dominant tracer at lowest [Fe/H] and earliest times in the
Universe.

Impetus for this possibility comes from the extreme dependence of
cooling at low metallicites and low temperatures on carbon and oxygen
(e.g., \citealp{dalgarno72}), and the suggestion by \citet{bromm03}
and \citet{frebel07b} that cooling through the fine-structure lines of
C\,II and O\,I played the major role in the collapse and fragmentation
of gas clouds in the early Universe, to produce the low mass stars we
observe today.  That is to say, the C-rich stars of
Table~\ref{tab:basic_cempno} are the survivors of the earliest times,
and objects that did not have C/Fe enhancements may no longer exist
among the most Fe-poor stars because (i) the first generations of low
C/Fe objects had a top-heavy IMF (i.e., having few low-mass stars)
and/or (ii) these generations took longer to form and enrich the
material from which later generations formed -- by which time the
C/Fe-enhanced stars had contributed significantly to the Fe abundance
of the Universe.

\item {\it From a limited data set of some 13 C-rich stars in
  Table~\ref{tab:basic_cempno}, only one exhibits evidence for radial
  velocity variations greater than 3 {\kms}}

There is currently little observational support for a universal binary
production of C-rich stars with [Fe/H] $<$ --3.0 (i.e., CEMP-no and
hyper metal-poor stars), such as exists for the CH stars and the
CEMP-s subclass.  That said, Monte Carlo analysis shows that binary
systems having periods greater than $\sim25$ years are not precluded
by the bulk of the available data.  A point worth noting is that none
of the (binary) CH- and CEMP-s stars has a period in this range; all
stars in these classes have P $\la$ 12 years.  A second interesting
distinction between the CEMP-no and the CEMP-s stars is that CEMP-s
stars are found only for [Fe/H] $\ga-3$, while CEMP-no stars exist at
all metallicities [Fe/H] $\la$ --2.0.

We conclude that the available data offer no clear support for a
binarity-related explanation of the C-rich stars with [Fe/H]
$\la-3.0$.  More velocities, systematically collected and on a
timescale of decades, will be needed before a definitive statement
based on radial velocity measurements can be made concerning the role
of binarity in the production of C-rich stars.

\end{enumerate}

\subsection{On the Origins of the C-rich and C-normal Populations} 

Within the $\Lambda$CDM paradigm of the early Universe, as described
by \citet[and references therein]{bromm09}, we suppose that the first
stars formed in dark matter ``minihalos'' from material containing no
elements heavier than lithium; that the cooling was provided by
molecular hydrogen; and that the mass function of these first objects
was top-heavy relative to those observed today, and contained no
low-mass, long-lived stars that might be observed today\footnote{We
  recognize that the detail of the mass function of the first stars is
  the subject of ongoing investigation (see, e.g., \citealp{clark11}
  and \citealp{dopcke12}).  As emphasized to us by a referee, given
  current uncertainty, one may only conclude that the population
  contained ``no stars with lifetimes longer than the age of the
  Universe''.}. These are the so-called Population III.1 stars.  We
further suppose that some fraction of these objects produced large
amounts of carbon and oxygen, as described, for example, by some or
all of the stellar evolutionary models of the type described above --
the rotating 250 -- 300~M$_{\odot}$ models of \citet{fryer01}; the
``mixing and fallback'' models of \citet{umeda03,umeda05}; the
relativistic jet-induced explosion of \citet{tominaga07b}; and the
rapidly-rotating stars of \citet {meynet06, meynet10}.  We also expect
that some fraction of the Population III.1 stars did not produce large
amounts of carbon (as the result perhaps of canonical SNe explosions
without fallback, or slower rotation), but produced chemical abundance
patterns that were rather more solar-like in nature.  The ejecta from
all of these objects provided the chemical enrichment of the material
that later formed the second generations (Population III.2).  We
consider two possible scenarios.

\subsubsection{Two cooling channels}

Following \cite{bromm03} and \citet{frebel07b}, we assume that during
the subsequent star formation within the second generation, the
material with large enhancements of carbon and oxygen fragmented to
form low-mass, long-lived stars that are still observed today.  We
identify the C-rich population with stars formed from the
carbon-enriched material.  Support for this identification comes in
particular from the work of \citet{frebel07b}, who investigated the
degree of carbon and/or oxygen enhancement that was necessary to
produce the cooling and subsequent fragmentation of the first
low-mass, long-lived stars. They introduced the transition
discriminant $D_{\rm trans} (={\rm log} (10^{{\rm [C/H]}} + 0.3 \times
10^{{\rm [O/H]}}$)) and predicted that no metal-poor stars should
exist below the critical value $D_{\mbox{\scriptsize trans}}$ =
$-3.5\pm 0.2$.  Inspection of the ultra metal-poor region
  ([Fe/H] $<$ --4.0) in Figure 23 ($D_{\rm trans}$ vs. [Fe/H]) of
  \citet{frebel11} (an update of Figure 1 of \citealp{frebel07b})
  shows that, of the four stars with [Fe/H] $\la$ --4.0, there are three
  above this critical value (the C-rich stars {\hen}, {\hej}. and
  {\hea} in our Table~\ref{tab:basic_cempno}), and one (the C-normal
  {\caffau11}) below it.  That is to say, the data for all C-rich
  stars with [Fe/H] $<$ --4.0 are consistent with the Bromm et
  al. hypothesis, while the non-C-rich star requires a different
  mechanism, as first pointed out by \citet{caffau11}.

We note in passing that we are unaware of models of very massive
rotating stars (M $\ga100$~M$_{\odot}$) that predict the observed,
correlated enhancements of, e.g., Na, Mg, and Al, relative to C and O.
Thus, at least from a nucleosynthetic point of view, further
theoretical work is required to establish if these objects played a
role in the chemical enrichment of the C-rich stars.

The existence of the C-normal star {\caffau}, with [Fe/H] = --4.7 and
[C/Fe] $<$ +0.9 \citep{caffau11, caffau12}, suggests that a different
gas-cooling mechanism also existed at the earliest times.  It lies
beyond the scope of the present work to identify that process, and we
refer the reader to \citet[and references therein]{bromm09} for
discussion of possibilities and uncertainties.  Bearing in mind the
caveats in that work, we draw the reader's attention to the
dust-induced star formation hypothesis of \citet{schneider06}, which
enables ``fragmentation to solar or subsolar mass scales already at
metallicities Z$_{\rm Cr}$ = 10$^{-6}$Z$_{\odot}$''.  See also the
more recent discussion by \citet{schneider12a, schneider12b}.  We
conjecture here that there was a second cooling channel in the early
Universe, and that it and cooling by C/O-rich material played
commensurate roles in producing the C-normal and C-rich populations,
respectively, having [Fe/H] $\la$ --3.0, observed today.

\subsubsection{One cooling channel plus binarity}

An alternative suggestion is that all stars resulted from the second
channel discussed in the previous paragraph, and that the C-rich stars
acquired their surface carbon later from processes involving binary
systems similar to those that produced CH stars and CEMP-s stars, and
as described, e.g., by \citet{suda04}, in the context of the C-rich
hyper metal-poor star {\hen}.  While we currently find no evidence for
the existence of a large fraction of binaries among the C-rich stars,
more work is needed before the binary hypothesis may be rigorously
excluded on observational grounds.

\subsection{Comparison with the Chemical Abundances of High Redshift, {\it z} = 2 -- 6, Galaxies}

How do the abundances of the most metal-poor Galactic halo stars
compare with those of high redshift galaxies?  We conclude by
comparing the stellar abundances discussed here with results for
galaxies observed in quasar absorption line systems having redshifts
{\it z} $>$ 2, in particular the metal-poor damped Lyman-$\alpha$
(DLA) systems at lower redshifts 2 $<$ {\it z} $\la$ 4 (e.g.,
\citealp[and references therein]{cooke11b}) and the so-called sub-DLAs
over the range 4 $\la$ {\it z} $<$ 6 (e.g., \citealp[and references
  therein]{becker12}).

In the lower redshift regime, \citet{cooke11a, cooke11b} report column
densities for H, C, N, O, Al, Si, S, Ar, Cr, Fe, Ni, and hence
relative abundances of the form [X/Y] (in particular [X/H]), as
adopted in the present work.  Three points of comparison are worth
making: (i) in a sample of 21 objects with [Fe/H] $\la$ --2.0, the
three most metal-poor systems have [Fe/H] = --3.0, --3.2, and --3.5;
(ii) the ratios of C/O and O/Fe are consistent with values determined
for stars in the Galactic halo (when the [OI]~6300{\AA} line is
adopted in the stellar analyses); and (iii) one of the 10 systems with
C and Fe abundances has the composition of a CEMP star -- [Fe/H] =
--3.0 and [C/Fe] = +1.5. (We note that this result has been challenged
by \citealp[their Section~4]{becker12}.  See also
\citealp{carswell12}.) 

The results of \citet{becker12} for the sub-DLAs extend the dataset to
redshift {\it z} = 6.3, and provide abundance information for C, O,
Si, and Fe.  Unfortunately, no estimates are available for the
abundance of hydrogen because ``the Ly$\alpha$ at {\it z} $\ga5$ is
too highly absorbed to allow accurate HI column density measurement'',
and no estimates of [X/H] (in particular [Fe/H]) are available in this
regime.  \citet{becker12} supplement their new results with those of
others at lower redshift (including those of \citealp{cooke11b},
except for their system having the composition of a CEMP star) to
provide a sample over the redshift range {\it z} = 2 -- 6.  The
extremely important limitation for a comparison of this collective
data set with the stellar abundances discussed in the present paper is
that we do not know the metallicities, [X/H], for all of the
high-redshift sample: indeed, the lowest available iron abundance is
[Fe/H] = --3.5, at redshift z = 3.7.  That said, Becker et al. (in
their Figure 11) plot [C/O], [Si/O], [C/Si], [C/Fe], [O/Fe], and
[Si/Fe] as functions of redshift, where one sees no evidence for a
large variation in any of the relative abundances.  In particular, for
their four systems having C and O abundances over the range 4.7 $<$
{\it z} $<$ 6.3, they report mean values $\langle$[C/Fe]$\rangle$ =
+0.17 $\pm$ 0.07 and $\langle$[O/Fe]$\rangle$ =+0.50 $\pm$ 0.05,
respectively.  That is to say, the C and O abundance of sub-DLA
systems at the highest redshifts currently observed are the same as
those of ``normal'' non-carbon-enhanced Galactic halo stars.  In
comparison with the abundances of carbon in the most Fe-poor stars in
the Milky Way, \citet{becker12} suggest: ``If carbon-enhanced stars
fairly reflect their native ISM abundances, then these abundances are
no longer common by {\it z} $\sim$ 6.  This raises the intriguing
possibility that most carbon-enhanced stars were formed at even
earlier times [than the C and O observed in the sub-DLA systems].''
Their conjecture resonates with our suggestion above that the C-rich
stars were the first low-mass, long-lived, stars to form in the
Universe.

\section{SUMMARY}\label{sec:summary}

We have examined the chemical abundance patterns of 18 carbon-rich
stars having [C/Fe] $\ge$ +0.7 and [Fe/H] $<-$3.1 (16 CEMP-no stars
and two other stars with [Fe/H] $\sim$ --5.5 and [C/Fe] $\sim$ +4, but no
star from the CEMP-r, r/s and -s subclasses), based on
high-resolution, high $S/N$, 1D model-atmosphere analyses.  These
objects represent some 30\% of stars below this iron-abundance limit
for which carbon abundances or limits permit C-rich and C-normal
determinations.  These C-rich stars are also oxygen- and
nitrogen-rich, while a large fraction of them is strongly enhanced in
Na, Mg, and Al relative to Fe, and to a lesser degree in Si and Ca.
These chemical signatures are consistent with the admixing and
processing of material from H-burning and He-burning regions, as
achieved by nucleosynthesis in the zero-heavy-element models of
``mixing and fallback'' SNe (\citealt{umeda03,umeda05}); of rotating,
massive and intermediate mass stars (\citealt {meynet06, meynet10});
and of Type II SNe with relativistic jets \citep{tominaga07b}.

We suggest that the C-rich and C-normal populations below [Fe/H]
$\sim$ --3.1 result from two different gas-cooling channels in the
very early Universe, of material that formed the progenitors of the
two populations. In the first, cooling was provided by fine-structure
line transitions of C\,II and O\,I to form the C-rich population.  In
the second, the physical process, while not well-defined (perhaps
dust-induced cooling?), led to the C-normal group.  The available
radial velocity data offer little support for a binary origin of these
C-rich stars (at least with periods less than $\sim25$ years), and
more data are required before one could conclude that binarity is
necessary for an understanding of the C-rich population.

A comparison of the abundances of the most Fe-poor, C-rich stars with
those reported for high-redshift Damped Lyman-$\alpha$ and sub-DLA
systems in the range {\it z} = 2 -- 6 is consistent with the view that
the C-rich stars originated at even earlier times than material
observed to date in the DLA and sub-DLA systems.

\acknowledgments

We thank the referee for perceptive and constructive criticisms of the
manuscript, which led to significant improvements.
J.\ E.\ N., D.\ Y., M.\ S.\ B., and
M.\ A. gratefully acknowledge support from the Australian Research
Council (grants DP03042613, DP0663562, DP0984924 and FL110100012) for
studies of the Galaxy's most metal-poor stars and ultra-faint
satellite systems.  J.\ E.\ N. thanks G.\ D.\ Becker,
R.\ F.\ Carswell, R.\ Cooke, and M.\ Pettini for valuable discussions.
N.\ C. acknowledges financial support for this work through the
Global Networks program of Universit\"at Heidelberg and
Sonderforschungsbereich SFB 881 ``The Milky Way System'' (subproject
A4) of the German Research Foundation (DFG).
R.\ F.\ G.\ W. acknowledges partial support
from the US National Science Foundation through grants AST-0908326 and
CDI-1124403.
T.\ C.\ B. acknowledges partial funding of this work from grants PHY
02-16783 and PHY 08-22648: Physics Frontier Center/Joint Institute for
Nuclear Astrophysics (JINA), awarded by the U.S. National Science
Foundation.  
P.\ S.\ B. acknowledges support from the Royal Swedish Academy of Sciences
and the Swedish Research Council; he is a Royal Swedish Academy of
Sciences Research Fellow supported by a grant from the Knut and Alice
Wallenberg Foundation.  


\clearpage

\clearpage

\clearpage

\clearpage

\clearpage

\end{document}

%% file: tab1.tex
\begin{deluxetable*}{lccrrrrrrl}
\tabletypesize{\scriptsize}
\tablecolumns{10}
\tablewidth{0pt}
\tablecaption{\label{tab:basic_cempno} BASIC DATA\tablenotemark{a} FOR 23 C-RICH (CEMP-NO AND TWO HYPER METAL-POOR\tablenotemark{b}) STARS}
\tablehead{
\colhead {Star} &  {RA2000}   &  {Dec2000} &  {\teff}  &{\logg} &{[Fe/H]} & {[C/Fe]}& {[Sr/Fe]} & {[Ba/Fe]} & {Source\tablenotemark{c}} \\ 
\colhead  {(1)} &  {(2)}  &  {(3)} &   {(4)}   & {(5)}  &  {(6)}  & { (7)}  &   {(8)}   &  {(9)}  &   {(10)}     
  }

\startdata

HE~0057$-$5959                      &   00 59 54.0 &$-$59 43 29.9 & 5257 & 2.65 & $-$4.08 &    0.86                  &  $-$1.06 &  $-$0.46 & 1             \\
HE~0107$-$5240\tablenotemark{b}                      &   01 09 29.2 &$-$52 24 34.2 & 5100 & 2.20 & $-$5.54 &    3.85 & $<-$0.52 &  $<$0.82 & 1, 2, 3       \\ 
53327-2044-515\tablenotemark{d}     &   01 40 36.2 &  +23 44 58.1 & 5703 & 4.68 & $-$4.00 &    1.13                  &     1.09 &  $<$0.34 & 1             \\ 
53327-2044-515\tablenotemark{d}     &              &              & 5703 & 3.36 & $-$4.09 &    1.57                  &     0.53 & $<-$0.04 & 1             \\ 
HE~0146$-$1548                      &   01 48 34.7 &$-$15 33 24.4 & 4636 & 0.99 & $-$3.46 &    0.84                  &  $-$0.38 &  $-$0.71 & 1             \\ 
{\ito}                           &   02 26 49.7 &  +44 57 46.0 & 5510 & 3.70 & $-$3.68 &    1.31                  &  $-$0.26 &  $-$0.59 & 4             \\ 
HE~0557$-$4840                      &   05 58 39.3 &$-$48 39 56.8 & 4900 & 2.20 & $-$4.81 &    1.70                  & $<-$1.07 &  $<$0.03 & 1, 5, 6       \\ 
Segue~1-7                           &   10 08 14.4 &  +16 05 01.0 & 4960 & 1.90 & $-$3.52 &    2.30                  &  $-$1.39 & $<-$0.96 & 7             \\ 
HE~1012$-$1540                      &   10 14 53.5 &$-$15 55 53.2 & 5745 & 3.45 & $-$3.47 &    2.22                  &  $-$0.37 &  $-$0.25 & 1, 8          \\ 
HE~1150$-$0428                      &   11 53 06.6 &$-$04 45 03.4 & 5208 & 2.54 & $-$3.47 &    2.37                  &  $-$0.12 &  $-$0.48 & 1, 9         \\ 
HE~1201$-$1512\tablenotemark{d}     &   12 03 37.0 &$-$15 29 33.0 & 5725 & 4.67 & $-$3.86 &    1.14                  & $<-$0.87 &  $<$0.05 & 1             \\ 
HE~1201$-$1512\tablenotemark{d}     &              &              & 5725 & 3.39 & $-$3.92 &    1.60                  & $<-$1.27 & $<-$0.34 & 1             \\ 
HE~1300+0157                        &   13 02 56.2 &  +01 41 52.1 & 5529 & 3.25 & $-$3.75 &    1.31                  &  $-$1.36 & $<-$0.85 & 1, 8, 10      \\ 
BS~16929-005                        &   13 03 29.5 &  +33 51 09.1 & 5229 & 2.61 & $-$3.34 &    0.99                  &     0.54 &  $-$0.41 & 1, 11, 12     \\ 
HE~1327$-$2326\tablenotemark{b}                      &   13 30 06.0 &$-$23 41 49.7 & 6180 & 3.70 & $-$5.76 &    4.26 &     1.04 &  $<$1.46 & 1, 13         \\ 
HE~1506$-$0113                      &   15 09 14.3 &$-$01 24 56.6 & 5016 & 2.01 & $-$3.54 &    1.47                  &  $-$0.85 &  $-$0.80 & 1             \\ 
CS~22878-027                        &   16 37 35.9 &  +10 22 07.8 & 6319 & 4.41 & $-$2.51 &    0.86                  &  $-$0.26 & $<-$0.75 & 1, 12         \\ 
CS~29498$-$043                      &   21 03 52.1 &$-$29 42 50.2 & 4639 & 1.00 & $-$3.49 &    1.90                  &  $-$0.35 &  $-$0.45 & 1, 14, 15     \\ 
HE~2139$-$5432                      &   21 42 42.4 &$-$54 18 42.9 & 5416 & 3.04 & $-$4.02 &    2.59                  &  $-$0.55 & $<-$0.33 & 1             \\ 
HE~2142$-$5656                      &   21 46 20.4 &$-$56 42 19.1 & 4939 & 1.85 & $-$2.87 &    0.95                  &  $-$0.19 &  $-$0.63 & 1             \\ 
HE~2202$-$4831                      &   22 06 05.8 &$-$48 16 53.0 & 5331 & 2.95 & $-$2.78 &    2.41                  &  $-$0.85 &  $-$1.28 & 1             \\ 
CS~29502-092                        &   22 22 36.0 &$-$01 38 27.5 & 5074 & 2.21 & $-$2.99 &    0.96                  &  $-$0.15 &  $-$1.20 & 1, 12         \\ 
HE~2247$-$7400                      &   22 51 19.4 &$-$73 44 23.6 & 4829 & 1.56 & $-$2.87 &    0.70                  & $<-$0.15 &  $-$0.94 & 1             \\ 
CS~22949-037\tablenotemark{e}                        &   23 26 29.8 &$-$02 39 57.9 & 4958 & 1.84 & $-$3.97 &    1.06 &     0.55 &  $-$0.52 & 1, 16, 17, 18 \\ 
CS~22957-027\tablenotemark{f}                        &   23 59 13.1 &$-$03 53 48.2 & 5170 & 2.45 & $-$3.19 &    2.27 &  $-$0.86 &  $-$0.80 & 1, 9, 19 \\ 

\enddata 
\tablenotetext{a}{Abundances based on one-dimensional LTE model-atmosphere analyses} 
\tablenotetext{b}{The hyper metal-poor {\hen} and {\hea}, with [Fe/H] $\sim-5.5$, may not be classified as CEMP-no stars, since only weak upper limits can be placed on their [Ba/Fe] values.}
\tablenotetext{c}{References: 1 = \citet{yong12}; 2 = \citet{christlieb04}, 3 = \citet{bessell04}, 4 =
  \citet{ito09}, 5 = \citet{norris07}, 6 = \citet{norris12a}, 7 = \citet{norris10b}, 8 = \citet{cohen08}, 9 =
  \citet{cohen06}, 10 = \citet{frebel07}, 11 = \citet{honda04b}, 12 =
  \citet{lai08}, 13 = \citet{aoki06}, 14 = \citet{aoki02b}, 15 =
  \citet{aoki04}, 16 = \citet{cayrel04}, 17 = \citet{spite05}, 18 =
  \citet{francois07}, 19 = \citet{norris97}}
\tablenotetext{d}{Dwarf and subgiant abundances from \citet{yong12}}
\tablenotetext{e}{HE~2323$-$0256} 
\tablenotetext{f}{HE~2356$-$0410}

\end{deluxetable*}

%% file: tab2.tex
\begin{deluxetable*}{lccrrrrrrl}
\tabletypesize{\scriptsize}
\tablecolumns{10}
\tablewidth{0pt}
\tablecaption{\label{tab:basic_cemprs} BASIC DATA\tablenotemark{a} FOR 26 CEMP STARS WITH [BA/FE] $>$ 0.}
\tablehead{
\colhead {Star} &  {RA2000}   &  {Dec2000} &  {\teff}  &{\logg} &{[Fe/H]} & {[C/Fe]}& {[Sr/Fe]} & {[Ba/Fe]} & {Source\tablenotemark{b}} \\ 
\colhead  {(1)} &  {(2)}      &  {(3)}     &   {(4)}   & {(5)}  &  {(6)}  & { (7)}  &   {(8)}   &  {(9)}    &   {(10)}     
  }
\startdata

CS~31062-050       &   00 30 31.1 &$-$12 05 10.9 & 5607 & 3.49 & $-$2.28 &    2.00 &     0.91 &     2.30 & 1, 2          \\          
SDSS~0036-10       &   00 36 02.2 &$-$10 43 36.3 & 6479 & 4.31 & $-$2.60 &    2.32 &  $-$0.07 &     0.40 & 1, 3          \\ 
CS~29497-034       &   00 41 39.8 &$-$26 18 54.4 & 4983 & 1.96 & $-$3.00 &    2.72 &      ... &     2.28 & 1, 4          \\ 
CS~31062-012       &   00 44 03.6 &$-$13 55 25.9 & 6190 & 4.47 & $-$2.67 &    2.12 &     0.33 &     2.32 & 1, 2, 3       \\ 
HE~0143$-$0441     &   01 45 37.9 &$-$04 26 43.4 & 6276 & 3.84 & $-$2.32 &    1.82 &     1.09 &     2.42 & 1, 5, 6       \\          
HE~0206$-$1916     &   02 09 19.6 &$-$19 01 55.5 & 5073 & 2.23 & $-$2.52 &    2.10 &      ... &     1.99 & 1, 4          \\  
HE~0207$-$1423     &   02 10 00.7 &$-$14 09 11.1 & 5023 & 2.07 & $-$2.95 &    2.38 &     0.33 &     1.73 & 1             \\ 
HE~0336+0113       &   03 38 52.9 &  +01 23 07.8 & 5819 & 3.59 & $-$2.60 &    2.25 &     1.87 &     2.69 & 1, 6          \\ 
HE~0441$-$0652     &   04 43 29.9 &$-$06 46 53.5 & 4811 & 1.52 & $-$2.77 &    1.38 &      ... &     1.20 & 1, 4          \\ 
52972-1213-507     &   09 18 49.9 &  +37 44 26.8 & 6463 & 4.34 & $-$2.98 &    2.82 &     1.15 &     1.70 & 1             \\ 
SDSS~0924+40       &   09 24 01.9 &  +40 59 28.8 & 6196 & 3.77 & $-$2.68 &    2.72 &     0.77 &     1.73 & 1, 3          \\ 
HE~1005$-$1439     &   10 07 52.4 &$-$14 54 21.0 & 5202 & 2.55 & $-$3.09 &    2.48 &      ... &     1.17 & 1, 4          \\ 
HE~1031$-$0020     &   10 34 24.2 &$-$00 36 08.4 & 5043 & 2.13 & $-$2.79 &    1.63 &     0.52 &     1.61 & 1, 6          \\ 
HE~1319$-$1935     &   13 22 38.7 &$-$19 51 11.6 & 4691 & 1.27 & $-$2.22 &    1.45 &      ... &     1.68 & 1, 4          \\  
HE~1429$-$0551     &   14 32 31.3 &$-$06 05 00.3 & 4757 & 1.39 & $-$2.60 &    2.28 &      ... &     1.47 & 1, 4          \\   
CS~30301-015       &   15 08 56.8 &  +02 30 18.5 & 4889 & 1.73 & $-$2.73 &    1.60 &     0.30 &     1.45 & 1, 2          \\ 
HD~196944          &   20 40 46.1 &$-$06 47 50.6 & 5255 & 2.74 & $-$2.44 &    1.20 &     0.84 &     1.10 & 1, 2          \\          
CS~22880-074       &   20 46 03.2 &$-$20 59 14.2 & 5621 & 3.50 & $-$2.29 &    1.30 &     0.39 &     1.31 & 1, 2          \\          
SDSS~2047+00       &   20 47 28.8 &  +00 15 53.8 & 6383 & 4.36 & $-$2.36 &    2.00 &     1.03 &     1.70 & 1, 3          \\   
CS~22948$-$027     &   21 37 45.8 &$-$39 27 22.3 & 5011 & 2.06 & $-$2.45 &    2.12 &      ... &     2.45 & 1, 4          \\  
HE~2158$-$0348     &   22 00 40.0 &$-$03 34 12.2 & 5150 & 2.44 & $-$2.57 &    1.87 &     0.64 &     1.75 & 1, 6          \\ 
CS~22892-052       &   22 17 01.7 &$-$16 39 27.1 & 4825 & 1.54 & $-$3.03 &    0.90 &     0.68 &     1.01 & 1, 7, 8, 9    \\ 
HE~2221$-$0453     &   22 24 25.6 &$-$04 38 02.2 & 4430 & 0.73 & $-$2.00 &    1.83 &      ... &     1.76 & 1, 4          \\  
HE~2228$-$0706     &   22 31 24.5 &$-$06 50 51.2 & 5003 & 2.02 & $-$2.78 &    2.32 &      ... &     2.46 & 1, 4          \\  
CS~30338-089       &   23 15 50.0 &  +10 19 26.2 & 4886 & 1.72 & $-$2.78 &    2.06 &      ... &     2.30 & 1, 4          \\ 
HE~2330$-$0555     &   23 32 54.8 &$-$05 38 50.6 & 4867 & 1.65 & $-$2.98 &    2.09 &      ... &     1.17 & 1, 4          \\  
\enddata
\tablenotetext{a}{Abundances based on one-dimensional LTE model-atmosphere analyses}
\tablenotetext{b}{References: 1= \citet{yong12}, 2 = \citet{aoki02}, 3
  = \citet{aoki08}, 4 = \citet{aoki07}, 5 = \citet{cohen04}, 6 =
  \citet{cohen06}, 7 = \citet{cayrel04}, 8 = \citet{spite05}, 9 =
  \citet{francois07} }
\end{deluxetable*}

%% file: tab3.tex
\begin{deluxetable}{lccrr}
\tabletypesize{\scriptsize}
\tablecolumns{5}
\tablewidth{0pt}
\tablecaption{\label{tab:barklem} CEMP STARS FROM BARKLEM ET AL. (2005)}
\tablehead{
\colhead   {Object} &  {[Fe/H]} & {[C/Fe]} & {[Ba/Fe]} & {Class\tablenotemark{a}} \\
\colhead   {(1)}    &  {(2)}   &  {(3)}    & {(4)}     & {(5)}                   \\ 
}
\startdata
HE~0131$-$3953  & $-$2.68 &   2.45 &    2.20 & r,s \\
HE~0202$-$2204  & $-$1.98 &   1.16 &    1.41 & r,s \\
HE~0231$-$4016  & $-$2.08 &   1.36 &    1.47 & r,s \\
HE~0338$-$3945  & $-$2.41 &   2.07 &    2.41 & r,s \\ 
HE~0430$-$4404  & $-$2.08 &   1.44 &    1.62 & r,s \\
HE~1105+0027    & $-$2.42 &   2.00 &    2.45 & r,s \\
HE~1124$-$2335  & $-$2.93 &   0.86 & $-$1.06 & no  \\
HE~1135+0139    & $-$2.31 &   1.19 &    1.13 & r,s \\
HE~1245$-$1616  & $-$2.97 &   0.77 &    0.28 & r,s  \\
HE~1300$-$0641  & $-$3.14 &   1.29 & $-$0.77 & no  \\
HE~1300$-$2201  & $-$2.60 &   1.01 & $-$0.04 & no  \\
HE~1330$-$0354  & $-$2.29 &   1.05 & $-$0.47 & no  \\
HE~1337+0012    & $-$3.44 &   0.71 &    0.07 & no  \\
HE~1343$-$0640  & $-$1.90 &   0.77 &    0.70 & r,s \\
HE~1351$-$1049  & $-$3.45 &   1.55 &    0.13 & no  \\
HE~1430$-$1123  & $-$2.70 &   1.84 &    1.82 & r,s \\
HE~2150$-$0825  & $-$1.98 &   1.35 &    1.70 & r,s \\
HE~2156$-$3130  & $-$3.13 &   0.74 &    0.52 & r,s \\
HE~2227$-$4044  & $-$2.32 &   1.67 &    1.38 & r,s \\
HE~2240$-$0412  & $-$2.20 &   1.35 &    1.37 & r,s \\
\enddata
\tablenotetext{a}{r,s = \citet{beers05} subclasses r, r/s and s} 
\end{deluxetable}

%% file: tab4.tex
\begin{deluxetable*}{lrrrrrrrrrrrrrrrrl}
\tabletypesize{\tiny}
\tablecolumns{18}
\tablewidth{0pt}
\tablecaption{\label{tab:abundances} RELATIVE ABUNDANCES\tablenotemark{a} AND $^{12}$C/$^{13}$C FOR THE 23 C-RICH STARS OF TABLE~\ref{tab:basic_cempno}}
\tablehead{
\colhead {Star} &{[Fe/H]} & 
{C}\tablenotemark{b}  & 
{$^{12}$C/$^{13}$C}    &
{N}\tablenotemark{b}  & 
{O}\tablenotemark{b}  &
{Na}\tablenotemark{b} & 
{Mg}\tablenotemark{b} & 
{Al}\tablenotemark{b} & 
{Si}\tablenotemark{b} & 
{Ca}\tablenotemark{b} & 
{Sc}\tablenotemark{b} & 
{Ti}\tablenotemark{b} & 
{Cr}\tablenotemark{b} & 
{Mn}\tablenotemark{b} & 
{Co}\tablenotemark{b} & 
{Ni}\tablenotemark{b} & {Source\tablenotemark{c}} \\ 
\colhead {(1)}  &  {(2)}  &  {(3)}  &   {(4)}  & {(5)}    &  {(6)} & { (7)} &   {(8)}&  {(9)} &  {(10)}&  {(11)}&  {(12)}&  {(13)}&  {(14)}&  {(15)}&  {(16)}&  {(17)} &  {(18)}     
  }
\startdata
HE~0057$-$5959                   & $-$4.08  &  0.86  & $>$2 &  2.15   & $<$2.77  &  1.98   &  0.51   &   ...   &   ...  &  0.65  &  0.17  &  0.40  & $-$0.50  &   ...   &   ...  &  0.17   & 1  \\
HE~0107$-$5240                   & $-$5.54  &  3.85  &$>$50 &  2.43   &    2.30  &  1.11   &  0.26   &   ...   &   ...  &  0.12  &   ...  &  0.04  &   ...    &   ...   &   ...  &   ...   & 1  \\
53327-2044-515\tablenotemark{d}  & $-$4.05  &  1.35  & $>$2 &   ...   & $<$2.81  &  0.14   &  0.40   & $-$0.17 &   ...  &  0.19  &  0.12  &  0.27  &   ...    &   ...   &   ...  &  0.12   & 1  \\
HE~0146$-$1548                   & $-$3.46  &  0.84  &    4 &   ...   & $<$1.63  &  1.17   &  0.87   &  0.14   &  0.50  &  0.22  &   ...  &  0.18  & $-$0.38  & $-$0.59 &  0.30  &  0.05   & 1  \\
{\ito}                        & $-$3.68  &  1.31  &  ... &  0.32   &    1.59  &  0.27   &  0.52   & $-$0.57 &  0.41  &  0.27  &  0.43  &  0.31  & $-$0.44  & $-$1.22 &  0.48  &  0.04   & 1  \\
HE~0557$-$4840                   & $-$4.81  &  1.70  &  ... &$<$1.00  &    2.30  & $-$0.18 &  0.17   & $-$0.65 &   ...  &  0.17  &   ...  &  0.36  & $-$0.69  &   ...   &   ...  & $-$0.17 & 1  \\
Segue~1-7                        & $-$3.52  &  2.30  &$>$50 &  0.75   & $<$2.21  &  0.53   &  0.94   &  0.23   &  0.80  &  0.84  &   ...  &  0.65  & $-$0.26  & $-$0.56 &  0.37  & $-$0.55 & 1  \\
HE~1012$-$1540                   & $-$3.47  &  2.22  &  ... &  1.25   &    2.25  &  1.93   &  1.85   &  0.65   &  1.07  &  0.70  &   ...  &  0.06  & $-$0.24  & $-$0.51 &  0.23  & $-$0.22 & 1  \\
HE~1150$-$0428                   & $-$3.47  &  2.37  &    4 &  2.52   &     ...  &   ...   &  0.41   &   ...   &   ...  &  1.16  &   ...  &  0.73  & $-$0.56  &   ...   &   ...  &   ...   & 1  \\
HE~1201$-$1512\tablenotemark{d}  & $-$3.89  &  1.37  &$>$20 &$<$1.26  & $<$2.64  & $-$0.33 &  0.24   & $-$0.73 &   ...  &  0.06  &  0.11  &  0.12  & $-$0.49  & $-$0.58 &  0.82  &  0.17   & 1  \\
HE~1300+0157                     & $-$3.75  &  1.31  & $>$3 &$<$0.71  &    1.76  & $-$0.02 &  0.33   & $-$0.64 &  0.87  &  0.39  &  0.30  &  0.36  & $-$0.38  & $-$0.76 &  0.49  &  0.08   & 1  \\
BS~16929-005                     & $-$3.34  &  0.99  & $>$7 &  0.32   &     ...  &  0.03   &  0.30   & $-$0.72 &  0.38  &  0.34  &  0.01  &  0.40  & $-$0.35  & $-$0.78 &  0.28  &  0.07   & 1  \\
HE~1327$-$2326                   & $-$5.76  &  4.26  & $>$5 &  4.56   &    3.70  &  2.48   &  1.55   &  1.23   &   ...  &  0.29  &   ...  &  0.80  &   ...    &   ...   &   ...  &   ...   & 1, 2  \\
HE~1506$-$0113                   & $-$3.54  &  1.47  &$>$20 &  0.61   & $<$2.32  &  1.65   &  0.89   & $-$0.53 &  0.50  &  0.19  &   ...  &  0.44  & $-$0.15  & $-$0.32 &  0.48  &  0.38   & 1  \\
CS~22878-027                     & $-$2.51  &  0.86  &  ... &$<$1.06  &     ...  & $-$0.17 & $-$0.11 &   ...   &  0.07  &  0.07  &  0.02  &  0.30  &  0.02  & $-$0.34   &   ...  &  0.09   & 1  \\
CS~29498-043                     & $-$3.49  &  1.90  &    6 &  2.30   &    2.43  &  1.47   &  1.52   &  0.34   &  0.82  &  0.00  &   ...  &  0.12  & $-$0.23  &   ...   &   ...  &   ...   & 1  \\
HE~2139$-$5432                   & $-$4.02  &  2.59  &$>$15 &  2.08   &    3.15  &  2.15   &  1.61   &  0.36   &  1.00  &$-$0.02\tablenotemark{e} &   ...  &  0.31  &  0.34    &   ...   &  0.62  &  0.17   & 1  \\
HE~2142$-$5656                   & $-$2.87  &  0.95  &  ... &  0.54   &     ...  &  0.81   &  0.33   & $-$0.62 &  0.35  &  0.30  &   ...  &  0.18  & $-$0.19  & $-$0.63 &  0.18  & $-$0.29 & 1  \\
HE~2202$-$4831                   & $-$2.78  &  2.41  &  ... &   ...   &     ...  &  1.44   &  0.12   &   ...   &   ...  &  0.17  &   ...  &  0.46  &  0.08    &   ...   &  0.21  & $-$0.07 & 1  \\
CS~29502-092                     & $-$2.99  &  0.96  &   20 &  0.81   &    0.75  &   ...   &  0.28   & $-$0.68 &   ...  &  0.24  &  0.30  &  0.28  & $-$0.26  & $-$0.48 &  0.23  &  0.16   & 1  \\
HE~2247$-$7400                   & $-$2.87  &  0.70  &  ... &   ...   &     ...  &  0.82   &  0.33   &   ...   &  0.80  &  0.43  &   ...  &  0.13  & $-$0.11  &   ...   &   ...  &  0.53   & 1  \\
CS~22949-037                     & $-$3.97  &  1.06  &    4 &  2.16   &    1.98  &  2.10   &  1.38   &  0.02   &  0.77  &  0.39  &  0.29  &  0.45  & $-$0.37  & $-$0.87 &  0.37  & $-$0.10 & 1, 3  \\
CS~22957-027                     & $-$3.19  &  2.27  &    6 &  1.75   &     ...  &   ...   &  0.30   & $-$0.10 &   ...  &  0.45  &   ...  &  0.52  & $-$0.17  & $-$0.10 &  0.22  &   ...   & 1  \\
\enddata
\tablenotetext{a}{Based on one-dimensional LTE model-atmosphere analyses}
\tablenotetext{b}{Abundances relative to iron: C = [C/Fe], N = [N/Fe], etc; Ti = [TiII/Fe] }
\tablenotetext{c}{References: 1 = This work (Section~\ref{sec:c1213} and Table~\ref{tab:basic_cempno}, column (10)); 2 = \citet{frebel06b}; 3 = \citet{depagne02}}
\tablenotetext{d}{Averages abundances of dwarf and subgiant solutions}
\tablenotetext{e}{Computed here following \citet{yong12} for an equivalent width of 61 m{\AA}}
\end{deluxetable*}

%% file: tab5.tex
\begin{deluxetable*}{lrrrrlr}
\tabletypesize{\scriptsize}
\tablecolumns{7}
\tablewidth{0pt}
\tablecaption{\label{tab:radvel} RADIAL VELOCITY DATA FOR C-RICH STARS FROM TABLE 1}
\tablehead{
\colhead{Star} & {V$_{\rm rad}$} &   {N}  & {Range}   & {Span}   & {Sources\tablenotemark{a}} & {Dist.\tablenotemark{b}} \\ 
\colhead {}    & {(\kms)}       &   { }  & {(\kms)}  & {(days)} & {}     & {(pc)}                        \\ 
\colhead{(1)}  &    {(2)}       &  {(3)} &  {(4)}    &  {(5}    & {(6)}  & {(7)}      \\ 
  }
\startdata
HE~0057$-$5959    &    375.3 &    1   &    ... &     ... &  2             &  8360   \\
HE~0107$-$5240    &     44.3 &    3   &    0.6 &     373 &  3             &  9620   \\
53327-2044-515    & $-$193.5 &    1   &    ... &     ... &  2             &  3110\tablenotemark{c}   \\
HE~0146$-$1548    & $-$114.9 &    1   &    ... &     ... &  2             & 37770   \\
{\ito}            & $-$150.6 &   28   &    3.2 &    4982 &  4             & 160     \\
HE~0557$-$4840    &    212.1 &    5   &    1.2 &     668 &  5, 6          & 10490   \\
Segue~1-7         &    204.3 &    1   &    ... &     ... &  7             & 23000   \\
HE~1012$-$1540    &    225.6 &    2   &    1.5 &    1093 &  8             &  1660   \\
HE~1150$-$0428    &     46.6 &    1   &    ... &     ... &  9             &  5970   \\
HE~1201$-$1512    &    238.0 &    2   &    1.6 &     279 &  2             &  1610\tablenotemark{c}   \\
HE~1300+0157      &     74.3 &    5   &    1.9 &     688 &  10, 11, 8     &  2050   \\
BS~16929-005      &  $-$51.8 &    3   &    1.2 &     308 &  12            &  3130   \\
HE~1327$-$2326    &     63.8 &    4   &    0.7 &     383 &  13, 14        &  1190   \\
HE~1506$-$0113    & $-$137.1 &    1   &    ... &     ... &  2             &  9710   \\
CS~22878-027      &  $-$90.8 &    1   &    ... &     ... &  12            &   ...   \\
CS~29498-043      &  $-$32.5 &    2   &    0.1 &     685 &  15, 16        & 13660   \\
HE~2139$-$5432    &    114.4 &    2   &    2.1 &     259 &  2             &  4720   \\
HE~2142$-$5656    &    103.4 &    1   &    ... &     ... &  2             &   ...   \\
HE~2202$-$4831    &     56.2 &    1   &    ... &     ... &  2             &   ...   \\
CS~29502-092      &  $-$68.7 &    3   &    1.1 &    1137 &  12            &   ...   \\
HE~2247$-$7400    &      5.7 &    1   &    ... &     ... &  2             &   ...   \\
CS~22949-037      & $-$125.8 &    4   &    0.8 &    4845 &  8, 17, 18, 19 &  9060   \\
CS~22957-027      &  $-$72.8 &   17   &   20.0 &    3125 &  20            &  3530   \\
\enddata 

\tablenotetext{a}{References: 1 = \citet{aoki08}, 2 = \citet{norris12b}, 3 = \citet{bessell04}, 4 = \citet{carney03},
  5 = \citet{norris07}, 6 = \citet{norris12a}, 7 = \citet{norris10b}, 8
  = \citet{cohen08}, 9 = \citet{cohen06}, 10 = \citet{frebel07}, 11 =
  \citet{barklem05}, 12 = \citet{lai08}, 13 = \citet{aoki06}, 14 =
  \citet{frebel06b}, 15 = \citet{aoki02b}, 16 = \citet{aoki04}, 17 =
  \citet{mcwilliam95}, 18 = \citet{norris01}, 19 = \citet{depagne02},
  20 = \citet{preston01}}
\tablenotetext{b}{Distances for stars having [Fe/H] $<$ --3.1}
\tablenotetext{c}{Distances for low gravity solutions.  The
  corresponding values for the high gravity cases are 620 pc and 340
  pc for 53327-2044-515 and HE~1201$-$1512, respectively}

\end{deluxetable*}

%% file: tab6.tex
\begin{deluxetable*}{lrrrrrlr}
\tabletypesize{\scriptsize}
\tablecolumns{8}
\tablewidth{0pt}
\tablecaption{\label{tab:radvel_cnormal} BASIC DATA FOR 35 C-NORMAL STARS WITH [FE/H] $\le-3.1$}
\tablehead{
\colhead{Star}   &   {\teff}  & {\logg}  &{[Fe/H]}   & {[C/Fe]}   & {V$_{\rm rad}$} & {Sources\tablenotemark{a}} & Dist. \\ 
\colhead {}      &   {}       &  {}      &  { }      & {  }       & {(\kms)}       &       & {(pc)}               \\ 
\colhead{(1)}    &   {(2)}    &  {(3)}   &  {(4)}    & {(5}       & {(6)} & {(7)}          & {(8)}                \\ 
  }
\startdata
CD~$-$38 245  & 4857 & 1.54 & $-$4.15 & $<-$0.33 &    46 &  1 &  4310 \\ 
CS~22183-031  & 5202 & 2.54 & $-$3.17 &     0.42 &    12 &  2 &  3230 \\ 
HE~0132--2439 & 5249 & 2.63 & $-$3.79 &     0.62 &   289 &  3 &  5380 \\ 
CS~22189-009  & 4944 & 1.83 & $-$3.48 &     0.31 &  --20 &  1 &  8050 \\ 
CS~22963-004  & 5597 & 3.34 & $-$3.54 &     0.40 &   292 &  4 &  2800 \\ 
CS~22968-014  & 4864 & 1.60 & $-$3.58 &     0.25 &   159 &  1 &  8930 \\ 
CS~22172-002  & 4893 & 1.68 & $-$3.77 &     0.00 &   251 &  1 &  4820 \\ 
BS~16469-075  & 4919 & 1.78 & $-$3.25 &     0.21 &   333 &  2 &  6300 \\ 
SDSS 1029+1729\tablenotemark{b} & 5811 & 4.00 & $-$4.73 & $<$0.93 & --34 &  5 & 1270 \\ 
BS~16920-017  & 4851 & 1.58 & $-$3.40 & $<-$0.07 & --206 &  2 &  9630 \\ 
BS~16085-050  & 4910 & 1.76 & $-$3.16 & $<-$0.52 &  --75 &  2 &  3570 \\ 
BS~16076-006  & 5566 & 3.32 & $-$3.51 &     0.34 &   206 &  6 &  1440 \\ 
HE~1320--2952 & 5106 & 2.26 & $-$3.69 &  $<$0.52 &   390 &  7 &  4940 \\ 
BS~16467-062  & 5310 & 2.80 & $-$3.80 &     0.40 &   -91 &  1 &  3200 \\ 
HE~1347--1025 & 5206 & 2.52 & $-$3.71 &     0.15 &    49 &  3 &  6260 \\ 
HE~1356--0622 & 4953 & 1.85 & $-$3.63 & $<-$0.05 &    94 &  3 &  8810 \\ 
BS~16550-087  & 4754 & 1.32 & $-$3.54 &  $-$0.49 & --147 &  4 & 11390 \\ 
HE~1424--0241 & 5260 & 2.66 & $-$4.05 &  $<$0.63 &    60 &  3 &  6570 \\ 
BS~16477-003  & 4879 & 1.66 & $-$3.39 &     0.29 & --223 &  1 & 10410 \\ 
CS~30325-094  & 4948 & 1.85 & $-$3.35 &     0.00 & --158 &  1 &  3570 \\ 
CS~30312-059  & 4908 & 1.75 & $-$3.22 &     0.27 & --156 &  4 &  5141 \\ 
BS~16084-160  & 4727 & 1.27 & $-$3.20 &  $-$0.12 & --130 &  4 &  9140 \\ 
CS~22878-101  & 4796 & 1.44 & $-$3.31 &  $-$0.29 & --129 &  1 &  9430 \\ 
BS~16080-093  & 4945 & 1.85 & $-$3.23 & $<-$0.63 & --205 &  4 &  6420 \\ 
CS~22891-209  & 4699 & 1.18 & $-$3.32 &  $-$0.65 &    80 &  1 &  6040 \\ 
BD~$-$18 5550 & 4558 & 0.81 & $-$3.20 &  $-$0.02 & --125 &  1 &  2230 \\ 
CS~22885-096  & 4992 & 1.93 & $-$3.86 &     0.26 & --250 &  1 &  5200 \\ 
CS~30336-049  & 4725 & 1.19 & $-$4.10 &  $<$0.23 & --237 &  4, 7  & 14110 \\ 
CS~22897-008  & 4795 & 1.43 & $-$3.50 &     0.56 &   267 &  1 &  8780 \\ 
CS~22948-066  & 5077 & 2.20 & $-$3.20 &     0.00 & --171 &  1 &  4420 \\ 
CS~22956-050  & 4844 & 1.56 & $-$3.39 &     0.27 &     0 &  1 & 11920 \\ 
CS~22965-054  & 6137 & 3.68 & $-$3.10 &     0.62 & --283 &  4 &  2300 \\ 
CS~29502-042  & 5039 & 2.09 & $-$3.27 &     0.16 & --138 &  1 &  3470 \\ 
CS~22888-031  & 6241 & 4.47 & $-$3.31 &     0.38 & --125 &  6 &   940 \\ 
CS~22952-015  & 4824 & 1.50 & $-$3.44 &  $-$0.41 &  --18 &  1 &  7890 \\ 
\enddata 
\tablenotetext{a}{References: 1 = \citet{bonifacio09}, 2 =
  \citet{honda04a}, 3 = \citet{cohen08}, 4 = \citet{lai08}, 5 =
  \citet{caffau11}, 6 = \citet{bonifacio07}, 7 = \citet{norris12b}}
\tablenotetext{b}{{\caffau} \citep{caffau11}}
\end{deluxetable*}

%% file: tab7.tex
\begin{deluxetable}{rrrr}
\tabletypesize{\scriptsize}
\tablecolumns{7}
\tablewidth{0pt}
\tablecaption{\label{tab:kinematics} SYSTEMIC KINEMATIC DATA FOR 53 STARS WITH [FE/H] $<$ --3.1}
\tablehead{
\colhead{$\langle$[Fe/H]$\rangle$}   &   {V$_{\rm rot}$}  &  {$\sigma_{\rm los}$}  &{No.}   \\ 
\colhead {}        &   {(\kms)}        &  {(\kms)}             &  {}       \\ 
\colhead{(1)}      &   {(2)}           &  {(3)}                &  {(4)}    \\ 
  }
\startdata
--3.32  &  --22$\pm$ 50 &  122$\pm$ 17  & 26 \\
--4.01  & --119$\pm$ 64 &  138$\pm$ 19  & 27 \\
\enddata 
\end{deluxetable}